\documentclass[a4paper,twocolumn,10pt,conference]{IEEEtran}

\usepackage[cmex10]{amsmath} % cmex10 proposed by IEEEtran_HOWTO
\usepackage{amssymb,amsthm,bm}
\usepackage{mathtools}
\usepackage{bm}

\usepackage{tikz}
\usetikzlibrary{calc,positioning,fit,shapes}
\usepackage{pgfplots}
\pgfplotsset{compat=1.9}

\usetikzlibrary{external}   
\usepackage{calc}
\usepackage{cite}

\makeatletter
\let\MYcaption\@makecaption
\makeatother

\usepackage[font=footnotesize]{subcaption}

\makeatletter
\let\@makecaption\MYcaption
\makeatother

\newtheorem{theorem}{Theorem}
\newtheorem{lemma}{Lemma}

\newcommand{\ie}{{i.e.},\ }

\newcommand{\Dinner}{\Delta^{(\mathsf{in})}}
\newcommand{\Douter}{\Delta^{(\mathsf{out})}}
\newcommand{\dii}{\delta_{i,j}^{(\mathsf{in})}}
\newcommand{\dil}{\delta_{i,j}^{(\mathsf{in},\ell)}}
\newcommand{\dia}{\delta_{i,j}^{(\mathsf{in},1)}}
\newcommand{\dib}{\delta_{i,j}^{(\mathsf{in},2)}}
\newcommand{\dic}{\delta_{i,j}^{(\mathsf{in},3)}}
\newcommand{\did}{\delta_{i,j}^{(\mathsf{in},4)}}
\newcommand{\doo}{\delta_{i,j}^{(\mathsf{out})}}
\newcommand{\dol}{\delta_{i,j}^{(\mathsf{out},\ell)}}
\newcommand{\doa}{\delta_{i,j}^{(\mathsf{out},1)}}
\newcommand{\dob}{\delta_{i,j}^{(\mathsf{out},2)}}
\newcommand{\doc}{\delta_{i,j}^{(\mathsf{out},3)}}
\newcommand{\dod}{\delta_{i,j}^{(\mathsf{out},4)}}

\newcommand{\given}{\, |\,}
\newcommand{\with}{\, ;\,}

\begin{document}

\title{Delivery Latency Trade-Offs of Heterogeneous Contents in Fog Radio Access Networks}

\author{
\IEEEauthorblockN{
Jasper Goseling
}
\IEEEauthorblockA{
 Stochastic Operations Research,\\
 University of Twente, The Netherlands\\
 j.goseling@utwente.nl
}
\and
\IEEEauthorblockN{
Osvaldo Simeone
}
\IEEEauthorblockA{
CWiP, ECE Department \\
NJIT, NJ, USA\\
osvaldo.simeone@njit.edu
}
\and
\IEEEauthorblockN{
Petar Popovski
}
\IEEEauthorblockA{
 Department of Electronic Systems, \\
 Aalborg University, Denmark\\
petarp@es.aau.dk
}
}

\maketitle

%%%%%%%%%%%%%%%%%%%%%%%%%%%%%%%%%%%%%%%%%%%%%%%%%%%%%%
%
%
%
%%%%%%%%%%%%%%%%%%%%%%%%%%%%%%%%%%%%%%%%%%%%%%%%%%%%%%
\begin{abstract}
A Fog Radio Access Network (F-RAN) is a cellular wireless system that enables content delivery via the caching of popular content at edge nodes (ENs) and cloud processing. The existing information-theoretic analyses of F-RAN systems, and special cases thereof, make the assumption that all requests should be guaranteed the same delivery latency, which results in identical latency for all files in the content library. In practice, however, contents may have heterogeneous timeliness requirements depending on the applications that operate on them. Given per-EN cache capacity constraint, there exists a fundamental trade-off among the delivery latencies of different users' requests, since contents that are allocated more cache space generally enjoy lower delivery latencies. For the case with two ENs and two users, the optimal latency trade-off is characterized in the high-SNR regime in terms of the Normalized Delivery Time (NDT) metric. The main results are illustrated by numerical examples.
\end{abstract}


%%%%%%%%%%%%%%%%%%%%%%%%%%%%%%%%%%%%%%%%%%%%%%%%%%%%%%
%
% Introduction
%
%%%%%%%%%%%%%%%%%%%%%%%%%%%%%%%%%%%%%%%%%%%%%%%%%%%%%%
\section{Introduction} \label{sec:intro}
Fog networking is a novel paradigm in which computing, storage and communication functions are implemented at both cloud and edge nodes (ENs), such as base stations, of a wireless cellular system. As Fig. 1 shows,
content delivery can benefit from fog networking via edge caching (storing popular content at the ENs), as well as via cloud processing, which enables the delivery of content fetched from a central content library.

The information-theoretic analysis of edge caching in \cite{AliNiesen15,nad016fundamental,cao2016fundamental,hachem2016degrees} and of more general fog-assisted wireless networks, or Fog Radio Access Networks (F-RANs), in \cite{TandonSimeone16,sengupta2016cloud} makes the assumption that all files in the content library have the same timeliness constraint. Under this assumption, caching schemes in which all contents are allocated the same fraction of the ENs' caches were proven to be optimal or near-optimal in \cite{TandonSimeone16,sengupta2016cloud}. In practice, however, contents may have heterogeneous latency requirements; e.g. video chunks may be buffered to reduce the delay constraints, while information feeding an Augmented Reality (AR) application has stricter latency requirements. Reducing the delivery latency of a content type generally requires allocating a larger fraction of the ENs' cache capacity to it, which in turn increases the delivery latency of other contents. 

As in \cite{AliNiesen15,nad016fundamental,cao2016fundamental,hachem2016degrees,TandonSimeone16,sengupta2016cloud}, in this paper, users are assumed to make simultaneous requests from a library of contents, which may be partially cached at the ENs during an offline caching phase. Unlike prior work, in which all request sets experience the same delivery coding latency, here we study the trade-offs among the latencies that are achievable across different request sets, when one allows arbitrary allocation of cache capacity at the ENs across files. Leveraging a fine-grained understanding of these trade-offs makes it possible to analyze individual content latency constraints, including the average latency for a content type under a probabilistic popularity model.

As in \cite{TandonSimeone16,sengupta2016cloud}, as well as in \cite{yi2016topological,cao2016fundamental}, delivery latencies are measured here in the high-Signal-to-Noise Radio (SNR) regime with respect to a reference interference-free system, yielding the performance metric of Normalized Delivery Time (NDT). In \cite{TandonSimeone16}, the minimum NDT was characterized for a system with two ENs and two users under the assumption that all users' requests should be guaranteed the same latency. Reference \cite{sengupta2016cloud} obtained upper and lower bounds that are within a multiplicative factor of 2 for any number of ENs and users.

%%%%%%%%%%%%%%%%%%%%%%%%%%%%%%%%%%%%%%%%%%%%%%%%%%%%%%
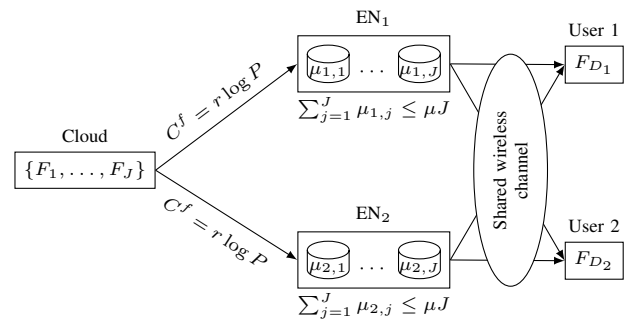
\begin{figure}
\centering
\begin{tikzpicture}[font=\scriptsize]

\node[
  draw,
  align=center
  ] (cloud) at (-1.5,0) {$\{F_1,\dots,F_J\}$};
\node[above=0mm of cloud] {Cloud};

\begin{scope}[xshift=23mm,yshift=13mm]
\node[
  draw,
  cylinder, shape border rotate=90, minimum width=5.5mm, minimum height=4mm
  ] (muaa) at (-0.6,0) {};
\node at (-0.6,0) {$\mu_{1,1}$};

\node at (0,0) {$\cdots$};

\node[
  draw,
  cylinder, shape border rotate=90, minimum width=5.5mm, minimum height=4mm
  ] (muab) at (0.6,0) {};
\node at (0.6,0) {$\mu_{1,J}$};

\node[draw,fit=(muaa) (muab)] (ENa) {};
\node[above=0mm of ENa] {EN$_1$};

\node at (0,-0.5) {$\sum_{j=1}^J\mu_{1,j}\leq \mu J$};
\end{scope}

\begin{scope}[xshift=23mm,yshift=-13mm]
\node[
  draw,
  cylinder, shape border rotate=90, minimum width=5.5mm, minimum height=4mm
  ] (muba) at (-0.6,0) {};
\node at (-0.6,0) {$\mu_{2,1}$};

\node at (0,0) {$\cdots$};

\node[
  draw,
  cylinder, shape border rotate=90, minimum width=5.5mm, minimum height=4mm
  ] (mubb) at (0.6,0) {};
\node at (0.6,0) {$\mu_{2,J}$};

\node[draw,fit=(muba) (mubb)] (ENb) {};
\node[above=0mm of ENb] {EN$_2$};

\node at (0,-0.5) {$\sum_{j=1}^J\mu_{2,j}\leq \mu J$};
\end{scope}

\node[draw] (usera) at (5.2,1.4) {$F_{D_1}$};
\node[above=0mm of usera] {User 1};

\node[draw] (userb) at (5.2,-1.2) {$F_{D_2}$};
\node[above=0mm of userb] {User 2};

\draw[-latex] (cloud.east) -- node[above,sloped] {$C^f=r\log P$} (ENa.west);
\draw[-latex] (cloud.east) -- node[below,sloped] {$C^f=r\log P$} (ENb.west);
\draw[-latex] (ENa.east) -- (usera.west);
\draw[-latex] (ENa.east) -- (userb.west);
\draw[-latex] (ENb.east) -- (usera.west);
\draw[-latex] (ENb.east) -- (userb.west);

\node[draw, fill=white,
  ellipse,
  align=center,
  inner xsep=3mm,
  rotate=90] (channel) at (4.1,0) {Shared wireless\\ channel};

\end{tikzpicture}
\caption{Illustration of the F-RAN system under study for $M=2$ and $K=2$.}
\label{fig:fran}
\vspace{-12pt}
\end{figure}
%%%%%%%%%%%%%%%%%%%%%%%%%%%%%%%%%%%%%%%%%%%%%%%%%%%%%%

Focusing on the special case with two ENs and two users as in \cite{TandonSimeone16}, the main contributions of this work are as follows: (\emph{i}) The performance metric of the \emph{NDT region} is introduced with the aim of analyzing the trade-off among the latencies achievable for individual users' requests under non-uniform cache partitions across files (Sec. II); (\emph{ii}) Novel achievable schemes are presented that yield an inner bound to the NDT region (Sec. IV); (\emph{iii}) Outer bounds on the NDT region are derived that conclusively characterize the NDT region (Sec. V); (\emph{iv}) Numerical results corroborate the analysis (Sec. VI). %%%%%%%%%%%%%%%%%%%%%%%%%%%%%
%

%%%%%%%%%%%%%%%%%%%%%%%%%%%%%%%%%%%%%%%%%%%%%%%%%%%%%%
%%%%%%%%%%%%%%%%%%%%%%%%%%%%%%%%%%%%%%%%%%%%%%%%%%%%%%
\section{System Model} \label{ssec:model}
\subsection{Model}
We consider an F-RAN architecture with $M$ edge nodes (ENs), which serve $K$ users over a shared wireless channel, see Fig. \ref{fig:fran}. As in prior works \cite{AliNiesen15,nad016fundamental,cao2016fundamental,hachem2016degrees,TandonSimeone16,sengupta2016cloud}, the system operates in two separate phases, namely an (offline) caching phase and an (online) delivery phase. In both phases, the content library $\mathcal{F}=\{F_1, \dots, F_J\}$ of $J\geq K$ files, where each file is of length $L$ bits, is fixed and static. The assumption of equally-sized files simplifies the treatment, as in prior work, and should be alleviated in future studies. In the caching phase, each EN $m$ can cache at most $\mu J L$ bits from the library, where $0\leq \mu \leq 1$ is referred to as the \emph{fractional cache capacity}. 

The delivery phase consists of an arbitrary number of slots. In any slot, each user $k$ requests a file $F_{D_k}$ in $\mathcal{F}$ with index $D_k \in [1:J]$. We let $D=[D_1,\dots,D_K]$ denote the vector of requested files in a slot. We make the assumption that the requested files are distinct, as e.g. in~\cite{pedarsani2016online}. In future work, we plan to alleviate this limitation. The main goal of this work is understanding the trade-offs achievable among the delivery latencies that are achievable for different request vectors $D$. As we discuss in Sec. \ref{sec:eval}, this fine-grained understanding of the trade-offs among the delivery latencies for different requests can be used to study individual latency requirements for different files under a given popularity distribution.

The channel from the ENs to the users is defined by:
\begin{equation}
Y_k = \sum_{m=1}^M H_{m,k}X_m + Z_k, \label{eq:rx_signal}
\end{equation}
which is a standard quasi-static model, where $X_m\in\mathbb{C}^{n^e}$ is a codeword of length $n^e$ symbols transmitted by the EN $m$; $H_{m,k}\in\mathbb{C}$ is the channel coefficient from EN $m$ to user $k$; $Z_k$ is complex Gaussian additive noise with unitary power, i.i.d. over time and users and also independent of the channel coefficients; and $Y_k\in\mathbb{C}^{n^e}$ is the received signal of length $n^e$ symbols by user $k$. The channel coefficients are realizations of continuous random variables, and are i.i.d. over ENs and users. Using the notation $[1:K]=\{1,\dots,K\}$, let $\mathcal{H}=(H_{m,k})_{m \in [1:M], k \in [1:K]}$ denote the channel state information (CSI), which is assumed to be known throughout the network, \ie at the cloud, the ENs and the users. 

The cloud has orthogonal fronthaul links to each of the ENs. Using the parametrization in \cite{sengupta2016cloud}, the capacity is measured in bits per symbol, where a \emph{symbol} is a channel use of the wireless channel. Furthermore, as in \cite{sengupta2016cloud}, the fronthaul capacity is written as $C^f = r\log(P)$, where $r$ is defined as the \emph{fronthaul rate} and $P$ is the (high) SNR of the wireless edge links. The fronthaul rate describes the ratio between the fronthaul capacity and the high-SNR capacity of each EN-to-user when used with no interference from other links. 

 In the \emph{caching phase}, EN $m$ stores an arbitrary function \begin{equation}
S_{m,j} = \pi_{m,j}^{c}(F_j)
\end{equation}
of each file $F_j, j \in [1:J]$,. We allow for an arbitrary partition of each EN's cache capacity. Denoting the entropy of the cached content for file $F_j$ at EN $m$ as $H(S_{m,j})=\mu_{m,j}L$, with $0 \leq \mu_{m,j} \leq 1$, we impose that the cache partition $\{\mu_{m,j}\}_{j \in [1:J]}$ satisfies the per-EN cache capacity constraint
\begin{equation} \label{eq:capacityconstraint}
\sum_{j=1}^J\mu_{m,j}\leq\mu J,
\end{equation}
for each $m$. We will refer to $\pi^c = (\pi^c_{m,j})_{m \in [1:M], j \in [1:J]}$ as the caching policy and to the matrix
\begin{equation} \label{eq:mumatrix}
\bm{\mu}=(\mu_{m,j})_{m\in[1:M], j \in [1:J]}
\end{equation}
as the \emph{cache partition matrix} for the given caching policy.

%\tikzexternaldisable
%%%%%%%%%%%%%%%%%%%%%%%%%%%%%%%%%%%%%%%%%%%%%%%%%%%%%%
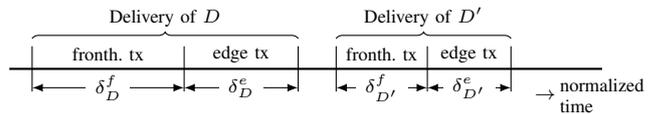
\begin{figure}
\centering
\begin{tikzpicture}[font=\scriptsize]

\draw[thick] (.7,0) -- node[at end, below,align=left,inner xsep=1pt] (ntime) {normalized\\ time} (8.5,0);
\node[left=0pt of ntime,inner xsep=0pt] {$\rightarrow$};

\draw (1,-.3) -- (1,.3);
\draw (3,-.3) -- (3,.3);
\draw (4.5,-.3) -- (4.5,.3);
\draw[latex-latex] (1,-.25) -- node[fill=white,inner ysep=0pt] {$\delta_D^f$} (3,-.25);
\path (1,.2) -- node[fill=white, inner ysep=0pt] {fronth.\ tx} (3,.2);
\draw[latex-latex] (3,-.25) -- node[fill=white, inner ysep=0pt] {$\delta_D^e$} (4.5,-.25);
\path (3,.2) -- node[fill=white, inner ysep=0pt] {edge tx} (4.5,.2);
\draw[decorate,decoration=brace] (1,.4) -- node[above=1pt] {Delivery of $D$} (4.5,.4);

\draw (5,-.3) -- (5,.3);
\draw (6.2,-.3) -- (6.2,.3);
\draw (7.3,-.3) -- (7.3,.3);
\draw[latex-latex] (5,-.25) -- node[fill=white,inner ysep=0pt] {$\delta_{D'}^f$} (6.2,-.25);
\path (5,.2) -- node[fill=white, inner ysep=0pt] {fronth.\ tx} (6.2,.2);
\draw[latex-latex] (6.2,-.25) -- node[fill=white, inner ysep=0pt] {$\delta_{D'}^e$} (7.3,-.25);
\path (6.2,.2) -- node[fill=white, inner ysep=0pt] {edge tx} (7.3,.2);
\draw[decorate,decoration=brace] (5,.4) -- node[above=1pt] {Delivery of $D'$} (7.3,.4);

\end{tikzpicture}
\caption{Delivery consists of a fronthaul transmission and an edge transmission phase. The length of these phases depends on the files that are requested.}
\label{fig:timeline}
\end{figure}
%%%%%%%%%%%%%%%%%%%%%%%%%%%%%%%%%%%%%%%%%%%%%%%%%%%%%%
%\tikzexternalenable

Each slot of the \emph{delivery phase} consists of two subsequent subslots (Fig.~\ref{fig:timeline}). In the \emph{first subslot}, the cloud sends information on the requested files to the ENs on the fronthaul links, while in the \emph{second subslot} the ENs use the shared wireless channel to transmit to the users.
To elaborate, in the first subslot, the cloud sends a message $U_m$ to the EN $m$ on the fronthaul as a function of the demand vector, the files and the CSI 
\begin{equation} \label{eq:pif}
U_m = \pi^f_m(D,\mathcal{F},\mathcal{H}).
\end{equation}
The first subslot has $n^f_D$ symbols, where we make explicit the dependence on the vector $D$, and the entropy of message $U_m$ must be bounded as $H(U_m) \leq C^{f}n^{f}_D$ in order to satisfy the fronthaul capacity constraints. We call $\pi^f=(\pi^f_1,\dots,\pi^f_M)$ the fronthaul policy.
In the second subslot, the ENs transmit a codeword $X_m$, of $n^e_D$ symbols, on the wireless channel as a function of the users' demand $D$, the cache content $S_m=\{S_{m,j}\}_{j \in [1:J]}$ of EN $m$, the fronthaul message $U_m$ to EN $m$ and the global CSI $\mathcal{H}$:
\begin{equation}
X_m = \pi^e_m(D, S_m, U_m, \mathcal{H}).
\end{equation}
We call $\pi^e=(\pi^e_1,\dots,\pi^e_M)$ the edge transmission policy. After receiving $Y_k$ in (\ref{eq:rx_signal}), user $k$ decodes the requested file as
\begin{equation}
\hat F_{D_k} = \pi^d_k(Y_k, D, \mathcal{H}),
\end{equation}
and we let $\pi^d=(\pi^d_1,\dots,\pi^d_K)$ denote the decoding policy. The error probability of a policy $\pi=(\pi^c, \pi^f, \pi^e, \pi^d)$ is defined as the worst-case error probability across requests and users
\begin{equation}
P_e = \max_{D} \max_{k\in[1:K]} P(\hat F_{k} \neq F_{D_k}).
\end{equation} A sequence of policies, parametrized by $L$ and $P$, is defined as \emph{feasible} if it satisfies the limit $\lim_{P\to\infty}\lim_{L\to\infty} P_e=0$. 

%%%%%%%%%%%%%%%%%%%%%%%%%%%%%%%%%%%%%%%%%%%%%%%%%%%%%%
%%%%%%%%%%%%%%%%%%%%%%%%%%%%%%%%%%%%%%%%%%%%%%%%%%%%%%
\subsection{Problem Statement} \label{ssec:problem}
For any sequence of feasible policies $\pi$ parametrized by $L$ and $P$, we now define the high-SNR delivery time metric for each demand vector $D$. To this end, we introduce the normalized durations of the first and second subslots in the given transmission interval as
\begin{equation}  \label{eq:NDT_components1}
\delta^x_D = \lim_{P\to\infty} \lim_{L\to\infty} \frac{n^x_D}{L/\log P},
\end{equation}
where $x=f$ for the first subslot (fronthaul transmission) and $x=e$ for the second subslot (edge transmission). In (\ref{eq:NDT_components1}), the subslot durations are normalized by the high-SNR delivery time of a reference system in which each user is served on an interference-free dedicated channel by an EN, namely $L/\log P$. Note, in fact, that an interference-free channel has a high-SNR capacity of $\log P$ (see also \cite{sengupta2016cloud} for additional discussion). We refer to $\delta^f_D$ and $\delta^e_D$ as the \emph{fronthaul and edge NDTs}, respectively, for request $D$. The overall NDT for request $D$ is hence given by $\delta_D = \delta^e_D + \delta^f_D$.

We are interested in characterizing the region $\Delta^{*}(\mu,r)$ of all achievable NDT tuples $\delta = (\delta_D)_{D\in\mathcal{D}}$ under the per-EN capacity constraint~\eqref{eq:capacityconstraint}, which we refer to as \emph{NDT region}. We impose that the same NDT be achieved $\delta_D$ be achieved for all permutations of the vector $D$. This allows us to obtain a characterization that depends only on the subset of files that are requested. Henceforth, with a slight abuse of notation, $D$ represents a subset of $[1:J]$. As a result, the NDT region is contained in the positive orthant of $\mathbb{R}^{J\choose K}$. 

To study the NDT region $\Delta^*(\mu, r)$ it is convenient to analyze also the region $\Delta^*(\bm{\mu},r)$ of all NDT tuples that are achievable with a given cache partition matrix $\bm{\mu}$ in (\ref{eq:mumatrix}).
By definition, we have
 \begin{equation} \label{eq:ndtasunion}
 \Delta^*(\mu,r) = \bigcup_{\bm{\mu}:\mathrm{ } \sum_{j=1}^J \mu_{m,j} \leq\mu, \mathrm{ }\forall m} \Delta^*(\bm{\mu}, r).
\end{equation}

As a first important observation is summarized in the following lemma.
\begin{lemma} \label{lem:convex}
	The NDT regions $\Delta^*(\mu,r)$ and $\Delta^*(\bm{\mu},r)$ are convex.
\end{lemma}
\emph{Sketch of proof}: The lemma follows from the fact that, given two NDT tuples $\delta$ and $\delta'$ that are achievable with cache partitions $\bm{\mu}$ and $\bm{\mu'}$, respectively, that satisfy~\eqref{eq:capacityconstraint}, the NDT tuple $\alpha \delta+(1-\alpha)\delta'$ (with entry-wise sum), for any $0 \leq \alpha \leq 1$, can also be achieved using cache sharing and file splitting for a cache partition $\alpha\bm{\mu}+(1-\alpha)\bm{\mu'}$ which satisfies~\eqref{eq:capacityconstraint}. A detailed proof can be found in Appendix~\ref{app:convex}. \qed

We finally note that the minimum NDT introduced in \cite{sengupta2016cloud}, corresponds to the minimum value $\delta$ in the NDT region $\Delta^*(\bm{\mu},r)$, with equal cache partition $\mu_{m,j}=\mu$, such that the equality $\delta=\delta_D$ holds for all request subsets $D$. In the rest of this paper, we focus on the special case $K=M=2$ and we write $\delta_D=\delta_{i,j}$ for any request subset $D=\{i, j\}$.

%\tikzexternaldisable
%%%%%%%%%%%%%%%%%%%%%%%%%%%%%%%%%%%%%%%%%%%%%%%%%%%%%%
\begin{figure}

\begin{subfigure}{\linewidth}
\centering
\begin{tikzpicture}[font=\scriptsize]

  \node (g1) at (-1.5,.3) {$G_1:$};
  \node (g2) at (-1.5,-.3) {$G_2:$};
  \node[right=0mm of g1, minimum width=15mm,draw] (g1bits) {$\nu L$ bits};
  \node[right=0mm of g2, minimum width=15mm,draw] (g2bits) {$\nu L$ bits};

  \node[fit=(g1) (g2) (g1bits) (g2bits), draw] (cloud) {};
  \node[above=0mm of cloud] {Cloud};

  \begin{scope}[xshift=33mm,yshift=8mm]
  \node[
    ] (muaa) at (-0.6,0) {$G_1$};

  \node[draw,fit=(muaa)] (ENa) {};
  \node[above=0mm of ENa] {EN$_1$};

  \end{scope}

  \begin{scope}[xshift=33mm,yshift=-8mm]
  \node[
    ] (muba) at (-0.6,0) {$G_2$};

  \node[draw,fit=(muba)] (ENb) {};
  \node[above=0mm of ENb] {EN$_2$};

  \end{scope}

  \draw[-latex,very thick] (cloud.east) -- node[above=5mm] {$\delta^f=\nu/r$} (ENa.west);
  \draw[-latex,very thick] (cloud.east) -- node[below,sloped] {} (ENb.west);

\end{tikzpicture}
\caption{Hard-transfer fronthauling (HT).}
\label{fig:existingA}
\end{subfigure}
%%%%%%%%%%%%%%%%%%%%%%%%%%%%%%%%%%%%%%%%%%%%%%%%%%%%%%

\medskip
%%%%%%%%%%%%%%%%%%%%%%%%%%%%%%%%%%%%%%%%%%%%%%%%%%%%%%
\begin{subfigure}{\linewidth}
\centering
\begin{tikzpicture}[font=\scriptsize]

\begin{scope}[xshift=0mm,yshift=10mm]
  \node (g1) at (0,0) {$G_1:$};
  \node[below=1mm of g1] (g2) {$G_2:$};
  \node[right=0mm of g1, minimum width=15mm,draw] (g1bits) {$\nu L$ bits};
  \node[right=0mm of g2, minimum width=15mm,draw] (g2bits) {$\nu L$ bits};
  \node[draw,fit=(g1) (g2) (g1bits) (g2bits)] (ENa) {};
\node[above=0mm of ENa] {EN$_1$};
\end{scope}

\begin{scope}[xshift=0mm,yshift=-10mm]
  \node (g1) at (0,0) {$G_1:$};
  \node[below=1mm of g1] (g2) {$G_2:$};
  \node[right=0mm of g1, minimum width=15mm,draw] (g1bits) {$\nu L$ bits};
  \node[right=0mm of g2, minimum width=15mm,draw] (g2bits) {$\nu L$ bits};
  \node[draw,fit=(g1) (g2) (g1bits) (g2bits)] (ENb) {};
\node[above=0mm of ENb] {EN$_2$};
\end{scope}

\node[ ] (useraa) at (4.2,0.7) {$G_1$};
\node[draw,fit=(useraa)] (usera) {};
\node[above=0mm of usera] {User 1};

\node[ ] (userbb) at (4.2,-1.25) {$G_2$};
\node[draw,fit=(userbb)] (userb) {};
\node[above=0mm of userb] {User 2};

\draw[-latex,very thick] (ENa.east) -- node[above=3mm] {$\delta^e=\nu$} (usera.west);
\draw[-latex,very thick] (ENa.east) -- (userb.west);
\draw[-latex,very thick] (ENb.east) -- (usera.west);
\draw[-latex,very thick] (ENb.east) -- (userb.west);

\end{tikzpicture}
\caption{Zero-forcing beamforming (ZF).}
\label{fig:existingB}
\end{subfigure}
%%%%%%%%%%%%%%%%%%%%%%%%%%%%%%%%%%%%%%%%%%%%%%%%%%%%%%

\medskip
%%%%%%%%%%%%%%%%%%%%%%%%%%%%%%%%%%%%%%%%%%%%%%%%%%%%%%
\begin{subfigure}{\linewidth}
\centering
\begin{tikzpicture}[font=\scriptsize]

  \node (g1) at (-1.5,.3) {$G_1:$};
  \node (g2) at (-1.5,-.3) {$G_2:$};
  \node[right=0mm of g1, minimum width=15mm,draw] (g1bits) {$\nu L$ bits};
  \node[right=0mm of g2, minimum width=15mm,draw] (g2bits) {$\nu L$ bits};

  \node[fit=(g1) (g2) (g1bits) (g2bits), draw] (cloud) {};
  \node[above=0mm of cloud] {Cloud};

  \begin{scope}[xshift=33mm,yshift=8mm]
  \node[minimum size=4mm
    ] (muaa) at (-0.6,0) {};

  \node[draw,fit=(muaa)] (ENa) {};
  \node[above=0mm of ENa] {EN$_1$};

  \end{scope}

  \begin{scope}[xshift=33mm,yshift=-8mm]
  \node[
     minimum size=4mm
    ] (muba) at (-0.6,0) {};

  \node[draw,fit=(muba)] (ENb) {};
  \node[above=0mm of ENb] {EN$_2$};

  \end{scope}

\node[ ] (useraa) at (5.2,.8) {$G_1$};
\node[draw,fit=(useraa)] (usera) {};
\node[above=0mm of usera] {User 1};

\node[ ] (userbb) at (5.2,-0.8) {$G_2$};
\node[draw,fit=(userbb)] (userb) {};
\node[above=0mm of userb] {User 2};

  \draw[-latex,very thick] (cloud.east) -- node[above=4mm] {$\delta^f=\nu/r$} (ENa.west);
  \draw[-latex,very thick] (cloud.east) -- node[below,sloped] {} (ENb.west);

\draw[-latex,very thick] (ENa.east) -- node[above=3mm] {$\delta^e=\nu$} (usera.west);
\draw[-latex,very thick] (ENa.east) -- (userb.west);
\draw[-latex,very thick] (ENb.east) -- (usera.west);
\draw[-latex,very thick] (ENb.east) -- (userb.west);

\end{tikzpicture}
\caption{Soft-transfer fronthauling with zero-forcing beamforming (ST+ZF).}
\label{fig:existingC}
\end{subfigure}
%%%%%%%%%%%%%%%%%%%%%%%%%%%%%%%%%%%%%%%%%%%%%%%%%%%%%%

\medskip
%%%%%%%%%%%%%%%%%%%%%%%%%%%%%%%%%%%%%%%%%%%%%%%%%%%%%%
\begin{subfigure}{\linewidth}
\centering
\begin{tikzpicture}[font=\scriptsize]

\begin{scope}[xshift=0mm,yshift=10mm]
  \node (g1) at (0,0) {$G_{1,1}:$};
  \node[below=1mm of g1] (g2) {$G_{1,2}:$};
  \node[right=0mm of g1, minimum width=15mm,draw] (g1bits) {$\nu L$ bits};
  \node[right=0mm of g2, minimum width=15mm,draw] (g2bits) {$\nu L$ bits};
  \node[draw,fit=(g1) (g2) (g1bits) (g2bits)] (ENa) {};
\node[above=0mm of ENa] {EN$_1$};
\end{scope}

\begin{scope}[xshift=0mm,yshift=-10mm]
  \node (g1) at (0,0) {$G_{2,1}:$};
  \node[below=1mm of g1] (g2) {$G_{2,2}:$};
  \node[right=0mm of g1, minimum width=15mm,draw] (g1bits) {$\nu L$ bits};
  \node[right=0mm of g2, minimum width=15mm,draw] (g2bits) {$\nu L$ bits};
  \node[draw,fit=(g1) (g2) (g1bits) (g2bits)] (ENb) {};
\node[above=0mm of ENb] {EN$_2$};
\end{scope}

\node[ ] (useraa) at (5.2,0.7) {$G_{1,1}$, $G_{2,1}$};
\node[draw,fit=(useraa)] (usera) {};
\node[above=0mm of usera] {User 1};

\node[ ] (userbb) at (5.2,-1.25) {$G_{1,2}$, $G_{2,2}$};
\node[draw,fit=(userbb)] (userb) {};
\node[above=0mm of userb] {User 2};

\draw[-latex,very thick] (ENa.east) -- node[above=3mm] {$\delta^e=3\nu$} (usera.west);
\draw[-latex,very thick] (ENa.east) -- (userb.west);
\draw[-latex,very thick] (ENb.east) -- (usera.west);
\draw[-latex,very thick] (ENb.east) -- (userb.west);

\end{tikzpicture}
\caption{X-channel interference alignment (X-IA).}
\label{fig:existingD}
\end{subfigure}
\caption{Illustration of constituent delivery strategies.}
\label{fig:existing}
\end{figure}
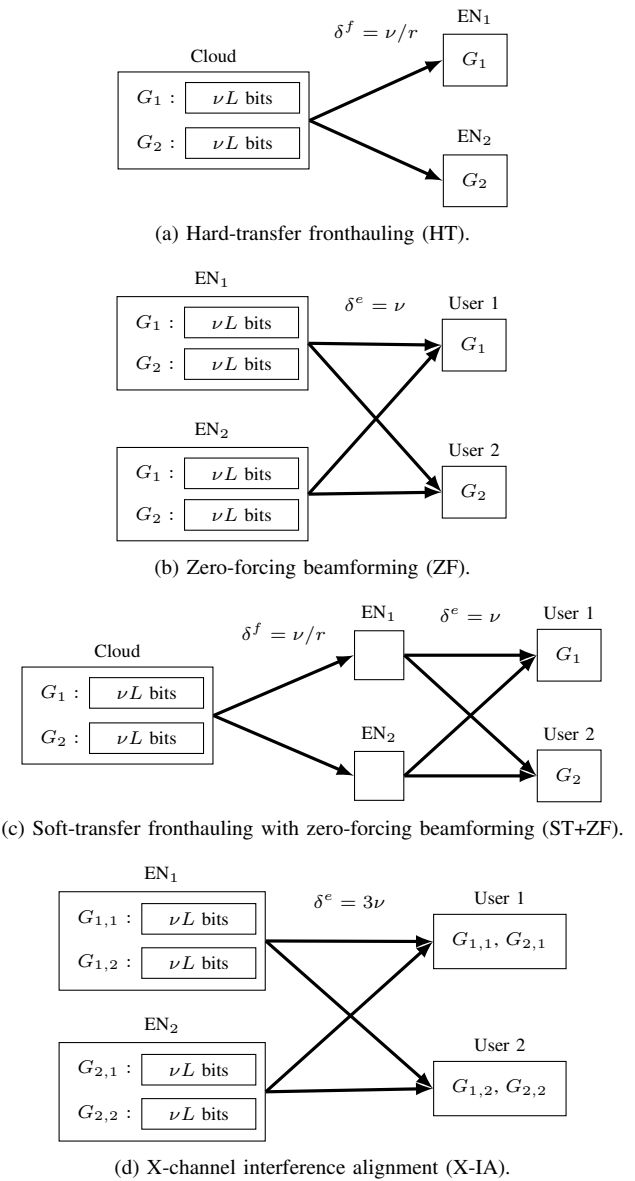
%%%%%%%%%%%%%%%%%%%%%%%%%%%%%%%%%%%%%%%%%%%%%%%%%%%%%%
%\tikzexternalenable

%\tikzexternaldisable
%%%%%%%%%%%%%%%%%%%%%%%%%%%%%%%%%%%%%%%%%%%%%%%%%%%%%%
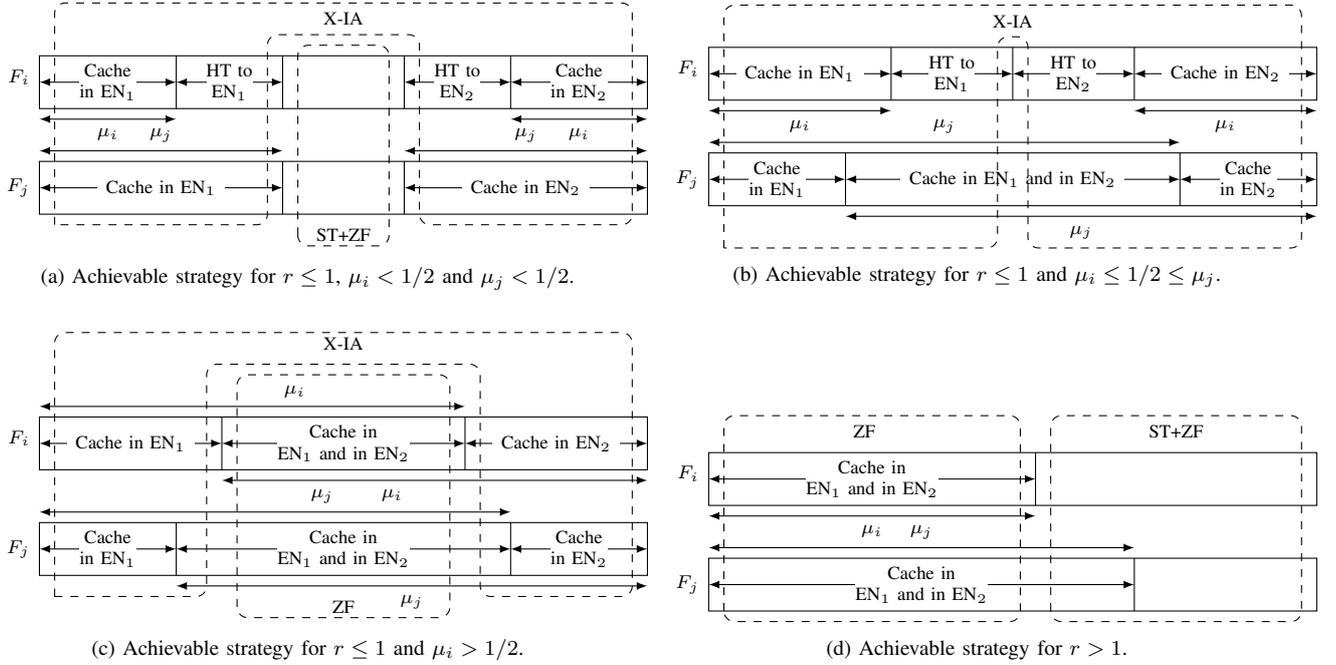
\begin{figure*}

\hfill
\begin{subfigure}{.45\linewidth}
\centering
\begin{tikzpicture}[font=\scriptsize,yscale=0.7]

\draw (0,2) rectangle (8,3);
\node[anchor=east] at (0,2.6) {$F_i$};
\draw (1.8,2) -- (1.8,3);
\draw (3.2,2) -- (3.2,3);
\draw[latex-latex] (0,1.8) -- node[below] {$\mu_i$} (1.8,1.8);
\draw[latex-latex] (0,2.5) -- node[fill=white,align=center,inner sep=0pt] {Cache\\ in EN$_1$} (1.8,2.5);
\draw[latex-latex] (1.8,2.5) -- node[fill=white,align=center,inner sep=0pt] {HT to\\ EN$_1$} (3.2,2.5);
\draw (6.2,2) -- (6.2,3);
\draw (4.8,2) -- (4.8,3);
\draw[latex-latex] (6.2,1.8) -- node[below,pos=0.5] {$\mu_i$} (8,1.8);
\draw[latex-latex] (6.2,2.5) -- node[fill=white,align=center,inner sep=0pt] {Cache\\ in EN$_2$} (8,2.5);
\draw[latex-latex] (4.8,2.5) -- node[fill=white,align=center,inner sep=0pt] {HT to\\ EN$_2$} (6.2,2.5);

\draw (0,0) rectangle (8,1);
\node[anchor=east] at (0,0.5) {$F_j$};
\draw (3.2,0) -- (3.2,1);
\draw[latex-latex] (0,1.2) -- node[above] {$\mu_j$} (3.2,1.2);
\draw[latex-latex] (0,0.5) -- node[fill=white] {Cache in EN$_1$} (3.2,0.5);
\draw (4.8,0) -- (4.8,1);
\draw[latex-latex] (4.8,1.2) -- node[above,pos=0.5] {$\mu_j$} (8,1.2);
\draw[latex-latex] (4.8,0.5) -- node[fill=white] {Cache in EN$_2$} (8,0.5);

\draw[rounded corners,dashed] (0.2,-0.2) -- (0.2,4) -- (7.8,4) -- (7.8,-0.2) -- (5,-0.2) -- (5,3.4) -- (3,3.4) -- (3,-.2) -- (0.2,-0.2);
\node at (4,3.7) {X-IA};

\draw[rounded corners,dashed] (3.4,-.6) rectangle (4.6,3.2);
\node at (4,-.4) {ST+ZF};

\end{tikzpicture}
\caption{Achievable strategy for $r\leq 1$, $\mu_i<1/2$ and $\mu_j<1/2$.}
\label{fig:ach1}
\end{subfigure}
\hfill
%\vspace{5mm}
%\medskip
%
%
\begin{subfigure}{.45\linewidth}
\centering
\begin{tikzpicture}[font=\scriptsize,yscale=0.7]

\draw (0,2) rectangle (8,3);
\node[anchor=east] at (0,2.6) {$F_i$};
\draw (2.4,2) -- (2.4,3);
\draw[latex-latex] (0,1.8) -- node[below] {$\mu_i$} (2.4,1.8);
\draw[latex-latex] (0,2.5) -- node[fill=white] {Cache in EN$_1$} (2.4,2.5);
\draw (4,2) -- (4,3);
\draw[latex-latex] (2.4,2.5) -- node[fill=white,align=center,inner sep=0pt] {HT to\\ EN$_1$} (4,2.5);
\draw (5.6,2) -- (5.6,3);
\draw[latex-latex] (4,2.5) -- node[fill=white,align=center,inner sep=0pt] {HT to\\ EN$_2$} (5.6,2.5);
\draw[latex-latex] (5.6,1.8) -- node[below,pos=0.5] {$\mu_i$} (8,1.8);
\draw[latex-latex] (5.6,2.5) -- node[fill=white] {Cache in EN$_2$} (8,2.5);

\draw (0,0) rectangle (8,1);
\node[anchor=east] at (0,0.5) {$F_j$};
\draw (1.8,0) -- (1.8,1);
\draw[latex-latex] (0,1.2) -- node[above] {$\mu_j$} (6.2,1.2);
\draw[latex-latex] (0,0.5) -- node[fill=white,align=center,inner sep=0pt] {Cache\\ in EN$_1$} (1.8,0.5);
\draw[latex-latex] (1.8,0.5) -- node[fill=white,align=center] {Cache in EN$_1$ and in EN$_2$} (6.2,0.5);
\draw (6.2,0) -- (6.2,1);
\draw[latex-latex] (1.8,-0.2) -- node[below] {$\mu_j$} (8,-0.2);
\draw[latex-latex] (6.2,0.5) -- node[fill=white,align=center,inner sep=0pt] {Cache\\ in EN$_2$} (8,0.5);

\draw[rounded corners,dashed] (0.2,-0.8) -- (0.2,3.8) -- (7.8,3.8) -- (7.8,-0.8) -- (4.2,-0.8) -- (4.2,3.2) -- (3.8,3.2) -- (3.8,-.8) -- (0.2,-0.8);
\node at (4,3.5) {X-IA};

\end{tikzpicture}
\caption{Achievable strategy for $r\leq 1$ and $\mu_i\leq 1/2\leq\mu_j$.}
\label{fig:ach2}
\end{subfigure}
\hfill{}

\vspace{5mm}

\hfill
\begin{subfigure}[b]{.45\linewidth}
\centering
\begin{tikzpicture}[font=\scriptsize,yscale=0.7]

\draw (0,2) rectangle (8,3);
\node[anchor=east] at (0,2.6) {$F_i$};
\draw (2.4,2) -- (2.4,3);
\draw[latex-latex] (0,3.2) -- node[above,pos=.6] {$\mu_i$} (5.6,3.2);
\draw[latex-latex] (0,2.5) -- node[fill=white] {Cache in EN$_1$} (2.4,2.5);
\draw[latex-latex] (2.4,2.5) -- node[fill=white,align=center,inner sep=0pt] {Cache in\\ EN$_1$ and in EN$_2$} (5.6,2.5);
\draw (5.6,2) -- (5.6,3);
\draw[latex-latex] (2.4,1.8) -- node[below,pos=0.4] {$\mu_i$} (8,1.8);
\draw[latex-latex] (5.6,2.5) -- node[fill=white] {Cache in EN$_2$} (8,2.5);

\draw (0,0) rectangle (8,1);
\node[anchor=east] at (0,0.5) {$F_j$};
\draw (1.8,0) -- (1.8,1);
\draw[latex-latex] (0,1.2) -- node[above,pos=.6] {$\mu_j$} (6.2,1.2);
\draw[latex-latex] (0,0.5) -- node[fill=white,align=center,inner sep=0pt] {Cache\\ in EN$_1$} (1.8,0.5);
\draw[latex-latex] (1.8,0.5) -- node[fill=white,align=center,inner sep=0pt] {Cache in\\ EN$_1$ and in EN$_2$} (6.2,0.5);
\draw (6.2,0) -- (6.2,1);
\draw[latex-latex] (1.8,-0.2) -- node[below] {$\mu_j$} (8,-0.2);
\draw[latex-latex] (6.2,0.5) -- node[fill=white,align=center,inner sep=0pt] {Cache\\ in EN$_2$} (8,0.5);

\draw[rounded corners,dashed] (0.2,-0.4) -- (0.2,4.6) -- (7.8,4.6) -- (7.8,-0.4) -- (5.8,-0.4) -- (5.8,4) -- (2.2,4) -- (2.2,-.4) -- (0.2,-0.4);
\node at (4,4.4) {X-IA};

\draw[rounded corners,dashed] (2.6,-.8) rectangle (5.4,3.8);
\node at (4,-.6) {ZF};

\end{tikzpicture}
\caption{Achievable strategy for $r\leq 1$ and $\mu_i> 1/2$.}
\label{fig:ach3}
\end{subfigure}
\hfill
\begin{subfigure}[b]{.45\linewidth}
\centering
\begin{tikzpicture}[font=\scriptsize,yscale=0.7]

\draw (0,2) rectangle (8,3);
\node[anchor=east] at (0,2.6) {$F_i$};
\draw (4.3,2) -- (4.3,3);
\draw[latex-latex] (0,1.8) -- node[below,pos=.5] {$\mu_i$} (4.3,1.8);
\draw[latex-latex] (0,2.5) -- node[fill=white,align=center,inner sep=0pt] {Cache in\\ EN$_1$ and in EN$_2$} (4.3,2.5);

\draw (0,0) rectangle (8,1);
\node[anchor=east] at (0,0.5) {$F_j$};
\draw (5.6,0) -- (5.6,1);
\draw[latex-latex] (0,1.2) -- node[above,pos=.5] {$\mu_j$} (5.6,1.2);
\draw[latex-latex] (0,0.5) -- node[fill=white,align=center,inner sep=0pt] {Cache in\\ EN$_1$ and in EN$_2$} (5.6,0.5);

\draw[rounded corners,dashed] (0.2,-.2) rectangle (4.1,3.7);
\node at (2.05,3.4) {ZF};

\draw[rounded corners,dashed] (4.5,-.2) rectangle (7.8,3.7);
\node at (6.15,3.4) {ST+ZF};

\end{tikzpicture}
\caption{Achievable strategy for $r>1$.}
\label{fig:ach4}
\end{subfigure}
\hfill{}

\caption{Achievable strategies}
\end{figure*}
%%%%%%%%%%%%%%%%%%%%%%%%%%%%%%%%%%%%%%%%%%%%%%%%%%%%%%
%\tikzexternalenable

%%%%%%%%%%%%%%%%%%%%%%%%%%%%%%%%%%%%%%%%%%%%%%%%%%%%%%
%%%%%%%%%%%%%%%%%%%%%%%%%%%%%%%%%%%%%%%%%%%%%%%%%%%%%%
\section{Preliminaries} \label{ssec:existing}
Here we review delivery strategies, see Fig.~\ref{fig:existing}, for the fronthaul and edge channels from \cite{sengupta2016cloud}, which will be used as ingredients in the next section to propose a more general caching and delivery policy. (1) \emph{Hard-transfer fronthauling} (HT, Fig.~\ref{fig:existingA}): As shows, via the HT fronthaul delivery strategy, the cloud delivers a fraction $\nu$ of one of the requested files, say $G_1$, to EN 1 and a fraction of the other file $G_2$ to EN 2 on the respective fronthaul links. (2) \emph{Zero-forcing beamforming} (ZF, Fig.~\ref{fig:existingB}): If both ENs have both requested messages $G_1$ and $G_2$, or a fraction $\nu$ thereof, available in the respective caches, the edge delivery strategy of cooperative ZF beamforming can be carried out on this fraction to deliver $G_i$ to user $i$, yielding parallel interference-free channels to both users. (3) \emph{Soft-transfer fronthauling with zero-forcing beamforming} (ST+ZF, Fig.~\ref{fig:existingC}): With the fronthaul-edge delivery strategy, the cloud implements ZF beamforming and transmits the resulting baseband signals to the ENs in quantized form. The ENs simply forward the quantized signals over the shared wireless channel \cite{simeone2009downlink}. (4) \emph{X-channel interference alignment} (X-IA, Fig.~\ref{fig:existingD}): If the ENs cache different fractions $\nu$ of each requested file, the resulting channel model for the delivery for this fraction is an X-channel, for which interference alignment (IA) edge delivery strategies were presented in~\cite{maddah2008communication}.

\begin{lemma} \label{lem:existing}\cite{TandonSimeone16}
The following fronthaul and edge NDTs %\label{eq:NDT_components}
are achievable using the delivery strategies summarized above.

\noindent{\em HT:} Let $G_1$ and $G_2$ be messages of $\nu L$ bits that are available in the cloud. HT requires the fronthaul NDT $\delta^f = \frac{\nu}{r}$ to transmit $G_1$ to EN$_1$ and $G_2$ to EN$_2$.

\noindent{\em ZF:} Let both ENs have messages $G_1$ and $G_2$ of $\nu L$ bits available. ZF requires the edge NDT $\delta^e = \nu$ to transmit $G_1$ to user $1$ and $G_2$ to user $2$.

\noindent{\em ST+ZF:} Let $G_1$ and $G_2$ be messages of $\nu L$ bits that are available in the cloud. ST+ZF requires the fronthaul and edge NDTs $\delta^f = \frac{\nu}{r}\quad\text{and}\quad\delta^e = \nu$ to transmit $G_1$ to user $1$ and $G_2$ to user $2$.

\noindent{\em X-IA:} Let $G_{i,1}$ and $G_{i,2}$ be messages of $\nu L$ bits that are available at EN$_i$, $i=1,2$. X-IA requires the edge NDT $\delta^e = 3\nu$ to transmit $G_{1,1}$ and $G_{2,1}$ to user $1$ and  $G_{1,2}$ and $G_{2,2}$ to user $2$.
\end{lemma}

%%%%%%%%%%%%%%%%%%%%%%%%%%%%%%%%%%%%%%%%%%%%%%%%%%%%%%
%
% Inner Bound
%
%%%%%%%%%%%%%%%%%%%%%%%%%%%%%%%%%%%%%%%%%%%%%%%%%%%%%%
\section{Achievable NDT Region} \label{sec:inner}
In this section, we present achievable strategies that yield an inner bound on the NDT region $\Delta^{*}(\mu,r)$. To this end, we consider policies with cache partitions $\bm{\mu}$ such that the two ENs cache the same number of bits for each file, i.e., $\mu_{1,j}=\mu_{2,j}=\mu_j$. As a result, each file $F_j$ is generally allocated a different cache fraction $\mu_j$ at the ENs. We will show in the next section that this restriction comes with no loss of optimality.
\begin{theorem} \label{th:inner}
An inner bound on the NDT region is given by the inclusion \begin{equation} \Delta^*(\mu,r) \supseteq\Dinner(\mu,r) = \bigcup_{\substack{\bm{\mu}: \mathrm{ } \mu_{1,i}=\mu_{2,i}=\mu_i,\\ \sum_{j=1}^J\mu_{j}\leq\mu}} \Dinner(\bm{\mu}, r),\end{equation}where the region
\begin{equation}
\Dinner(\bm{\mu},r)= \left\{\delta_D \middle| \delta_{i,j}\geq\dii(\bm{\mu},r), \forall \{i,j\}\in\mathcal{D}\right\}
\end{equation}
is included in $\Delta^*(\bm{\mu},r)$, and we have
\begin{align} \label{eq:deltainner}
\dii(\bm{\mu},r) =
\begin{cases}
\dia(\bm{\mu},r),\quad &\text{if } r\leq 1, \mu_i<\frac{1}{2} \text{ and } \mu_j < \frac{1}{2}, \\
\dib(\bm{\mu},r),\quad &\text{if } r\leq 1 \text{ and }  \\
&\ \hspace{-6.6mm}(\mu_i\leq\frac{1}{2}\leq\mu_j, \text{ or } \mu_j\leq\frac{1}{2}\leq\mu_i), \\
\dic(\bm{\mu},r),\quad &\text{if } r\leq 1, \mu_i>\frac{1}{2} \text{ and } \mu_j>\frac{1}{2}, \\
\did(\bm{\mu},r),\quad &\text{if } r>1,
\end{cases}
\end{align}
with the definitions
\begin{align}
\dia(\bm{\mu},r) =&\ 1 + \frac{1}{r} - \left(\frac{1}{r} - 1\right)\max\{\mu_i, \mu_j\} \notag \\
& - \frac{1}{r}\min\{\mu_i, \mu_j\}, \\
\dib(\bm{\mu},r) =&\ \frac{3}{2} + \frac{1}{r}\left(\frac{1}{2}-\min\{\mu_i,\mu_j\}\right), \\
\dic(\bm{\mu},r) =&\ 2 - \min\{\mu_i,\mu_j\}, \\
\did(\bm{\mu},r) =&\ 1 + \frac{1}{r} - \frac{1}{r}\min\{\mu_i,\mu_j\}.
\end{align}
\end{theorem}
In the remainder of this section, we present the achievable strategies that yield the inner bound in the previous theorem at an intuitive level. The detailed proof of Theorem~\ref{th:inner} is given in Appendix~\ref{app:inner}. 
To this end, we will present two different caching policies for the cases $r\leq 1$ and $r>1$. Note that the caching policy cannot depend on the demand $D=\{i,j\}$, unlike the delivery policy. In the following, we set $\mu_i\leq \mu_j$ without loss of generality.

\noindent \emph{Caching policy for $r\leq 1$}: As seen in Figures~\ref{fig:ach1}--\ref{fig:ach3}, each file $F_i$ is cached so that the bits indexed by $1,\dots,\mu_i L$ are stored in EN$_1$ and bits $(1-\mu_i)L,\dots, L$ are stored in EN$_2$, \ie we minimize the overlap in the cached content in EN$_1$ and EN$_2$ by storing the first part of the file in EN$_1$ and the last part of the file in EN$_2$. If $\mu_i\geq 1/2$ some overlap will occur and some bits will be stored in both ENs.

\noindent \emph{Caching policy for $r>1$}: As seen in Figure~\ref{fig:ach4}, each file $F_i$ is cached so that bits $1,\dots,\mu_i L$ in both EN$_1$ and EN$_2$, \ie we cache only the first part of the file and we maximize the overlap between the content that is cached in the ENs.

\noindent \emph{Delivery strategy for $\mu_j<1/2$ and $r\leq 1$}: This case is illustrated in Figure~\ref{fig:ach1} and achieves $\dia(\bm{\mu},r)$. The delivery proceeds in three phases: a) we use HT to deliver bits $\mu_iL,\dots\mu_jL$ and $(1-\mu_j)L,\dots,(1-\mu_i)L$ of file $F_i$ to EN$_1$ and EN$_2$, respectively; b) we use X-IA to transmit bits $1,\dots,\mu_jL$ and $(1-\mu_j)L,\dots,L$ of the files from the ENs to the users; c) we use ST+ZF to transmit bits $\mu_jL,\dots,(1-\mu_j)L$ directly from the cloud. Note that the strategy in~\cite{sengupta2016cloud} does not require step a). In fact, interestingly, the optimal policy in~\cite{sengupta2016cloud} did not make any use of HT fronthauling. The optimality results presented in the next section demonstrate that, instead, when $\mu_i\neq\mu_j$, the joint use of both HT and ST are instrumental in achieving the optimal NDT performance.

\noindent \emph{Delivery strategy for $\mu_i\leq 1/2\leq\mu_j$ and $r\leq 1$}: This case is illustrated in Figure~\ref{fig:ach2} and achieves $\dib(\bm{\mu},r)$. The delivery proceeds in two phases: a) we use HT to deliver bits $\mu_iL,\dots L/2$ and $L/2,\dots,(1-\mu_i)L$ of file $F_i$ to EN$_1$ and EN$_2$, respectively; and b) we use X-IA to deliver both complete files from the ENs to the users. We remark that this scenario is not relevant for the special case from~\cite{sengupta2016cloud}. We emphasize the important role of HT for deriving an achievable strategy, which is used here but not in~\cite{sengupta2016cloud}, 

\noindent \emph{Delivery strategy for $\mu_i>1/2$ and $r\leq 1$}: This case is illustrated in Figure~\ref{fig:ach3} and achieves $\dic(\bm{\mu},r)$. The delivery proceeds in two phases: a)  we use X-IA to deliver bits $1,\dots,(1-\mu_i)L$ and $\mu_iL,\dots,L$ of the files from the ENs to the users; and b) we use ZF to transmit bits $(1-\mu_i)L,\dots,\mu_iL$ from the ENs to the users. Note that this strategy does not make use of the fronthaul during the delivery.

\noindent \emph{Delivery strategy for $r > 1$}: The first $\min\{\mu_i, \mu_j\}L$ bits of both files, which are stored at both ENs, are delivered using ZF. The remaining $(1-\min\{\mu_i,\mu_j\})L$ bits are delivered using ST+ZF. The strategy is illustrated in Fig.~\ref{fig:ach4} and achieves $\did(\bm{\mu},r)$.

%%%%%%%%%%%%%%%%%%%%%%%%%%%%%%%%%%%%%%%%%%%%%%%%%%%%%%
%
% Outer Bound
%
%%%%%%%%%%%%%%%%%%%%%%%%%%%%%%%%%%%%%%%%%%%%%%%%%%%%%%
\section{Characterization of the NDT Region} \label{sec:outer}
In this section, we present an outer bound on the NDT region $\Delta^*(\mu,r)$ and we prove that the inner bound from the previous section is in fact tight, hence characterizing the NDT region. The first result of this section provides an outer bound on the achievable NDT tuple region for a fixed, and generic, cache partition $\bm{\mu}$.
\begin{theorem} \label{th:outer} For any cache partition $\bm{\mu}$, we have the outer bound $\Delta^*(\bm{\mu},r)\subseteq\Douter(\bm{\mu}, r)$, where
\begin{equation}
\Douter(\bm{\mu}, r) = \left\{ \delta\ \middle|\ \delta_{i,j} \geq \doo(\bm{\mu},r), \forall \{i,j\}\in\mathcal{D}  \right\},
\end{equation}
with
\begin{align} \label{eq:deltaouter}
\doo(\bm{\mu},r) =
\begin{cases}
\max_{\ell=1,\dots,3}\left\{\dol(\bm{\mu},r)\right\},\quad &\text{if } r\leq 1, \\
\dod(\bm{\mu},r),\quad &\text{if } r>1,
\end{cases}
\end{align}
and the definitions
\begin{multline}
\doa(\bm{\mu},r) = 1 + \frac{1}{r} - \min\left\{\mu_{1,i}, \mu_{2,i}, \mu_{1,j}, \mu_{2,j}\right\} \\
-\frac{1}{2}\left(\frac{1}{r}-1\right)\left(  \mu_{1,i} +  \mu_{2,i} + \mu_{1,j} + \mu_{2,j}  \right), \label{eq:thconverseA}
\end{multline}
\begin{multline}
\dob(\bm{\mu},r) = \frac{3}{2} + \frac{1}{2r} - \min\left\{\mu_{1,i}, \mu_{2,i}, \mu_{1,j}, \mu_{2,j}\right\} \\
- \frac{1}{2}\left(\frac{1}{r}-1\right) \min\{\mu_{1,i} + \mu_{2,i}, \mu_{1,j} + \mu_{2,j}\}, \label{eq:thconverseB}
\end{multline}
\begin{equation}
\doc(\bm{\mu},r) = 2 - \min\left\{\mu_{1,i}, \mu_{2,i}, \mu_{1,j}, \mu_{2,j}\right\}, \label{eq:thconverseC}
\end{equation}
\begin{equation}
\dod(\bm{\mu},r) = 1 + \frac{1}{r} - \frac{1}{r}\min\left\{\mu_{1,i}, \mu_{2,i}, \mu_{1,j}, \mu_{2,j}\right\}. \label{eq:thconverseD}
\end{equation}

\end{theorem}
The proof of Theorem~\ref{th:outer} is given in Appendix~\ref{app:outer}. 
Using the outer bound in the previous theorem, we show that the inner bound from the previous section is tight. This result implies that, in order to exhaust the NDT region, it is sufficient to consider cache partitions in which $\mu_{1,j}=\mu_{2,j}$ for all files $j \in [1:J].$
\begin{theorem} \label{th:region} The NDT region is given as $\Delta^*(\mu,r) = \Dinner(\mu,r)$.
\end{theorem}
The proof of Theorem~\ref{th:region} is given in Appendix~\ref{app:region}.

%%%%%%%%%%%%%%%%%%%%%%%%%%%%%%%%%%%%%%%%%%%%%%%%%%%%%%
%
% Two Classes of Files
%
%%%%%%%%%%%%%%%%%%%%%%%%%%%%%%%%%%%%%%%%%%%%%%%%%%%%%%
\section{Numerical Example} \label{sec:eval}
Consider a set-up in which the set of popular files is partitioned into two disjoint classes as $\mathcal{F}=\mathcal{F}_{1}\cup\mathcal{F}_{2}$, where class $\mathcal{F}_{i}$ has $J_i$ files. We first illustrate the NDT region derived above and we then discuss how this can be used to obtain optimal trade-offs among the individual latencies of different files under a popularity distribution.

We first illustrate a slice of the NDT region in which we impose that the same NDT $\delta_{(i),(j)}$ be achieved for all subsets $D$ for which one file is in the class $\mathcal{F}_i$ and the other in $\mathcal{F}_j$. Recall that we assume that the requested files are distinct (even if requested from the same class). The considered slice of the NDT region is three-dimensional with axes given by $\delta_{(1),(1)}$, $\delta_{(2),(2)}$ and $\delta_{(1),(2)}$. To further reduce the dimensionality, we let $\delta_{(2),(2)}$ be arbitrary, so as to focus only on the plane $(\delta_{(1),(2)}, \delta_{(1),(1)})$. In order to evaluate the boundary of this slice of the NDT region, it can be argued that it is sufficient to consider cache partitions such as all files within the same class, which we denote as $\mu_{(1)}$ and $\mu_{(2)}$ for the files of class $1$ and $2$, respectively. With this choice, the cache capacity constraint~\eqref{eq:capacityconstraint} reduces to $J_1\mu_{(1)} + J_2\mu_{(2)} \leq \mu(J_1+J_2)$.

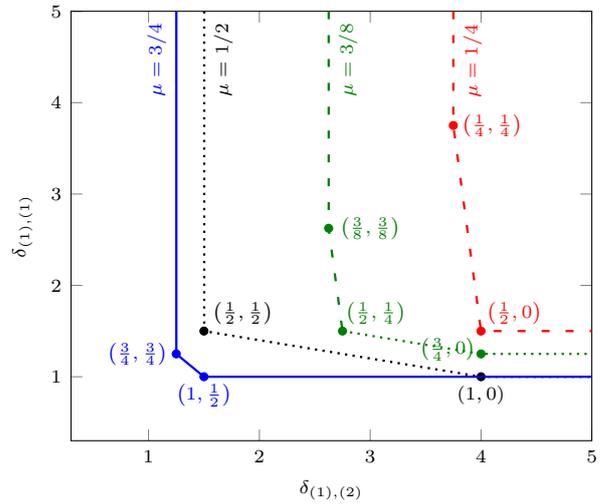
\begin{figure}
\centering
\begin{tikzpicture}
\tikzstyle{point}=[draw,fill,circle,minimum size=3pt,inner sep=0pt]
%\pgfplotsset{every axis legend/.append style={
%        at={(0.5,1.03)},
%        anchor=south}}
\begin{axis}[
	xlabel=$\delta_{(1),(2)}$,
	ylabel=$\delta_{(1),(1)}$,
	xmin=0.3,
	xmax=5,
	ymin=0.3,
	ymax=5,
	font=\scriptsize,
%	legend columns=2,
%	legend cell align=left,
%	legend style={
%          cells={anchor=west},
%          legend pos=outer north,
%         }
]
\addplot[
  line width=.3mm,color=red, loosely dashed,
%  mark=square*,mark repeat=50,mark phase=0,mark size=.5mm,mark options={solid}
  ]
table[
  header=false,x index=0,y index=1,
  ]
{matlab/two_classes2ext.csv}
node [pos=0.12,rotate=90,anchor=north] {$\mu=1/4$};
\node[point,red] (pa) at (axis cs: 4,1.5) {};
\node[red,align=center] at (axis cs: 4.3,1.7) {$\left(\frac{1}{2}, 0\right)$}; %{$\mu_{(1)}=1/2,$\\ $\mu_{(2)}=0$};
\node[point,red] (pb) at (axis cs: 3.75,3.75) {};
\node[red,align=center] at (axis cs: 4.1,3.75) {$\left(\frac{1}{4}, \frac{1}{4}\right)$}; %{$\mu_{(1)}=1/4,$\\ $\mu_{(2)}=1/4$};
%\addlegendentry{$\mu=1/4$};
%
%
\addplot[
  line width=.3mm,color=green!50!black, loosely dashed,
%  mark=square*,mark repeat=50,mark phase=0,mark size=.5mm,mark options={solid}
  ]
table[
  header=false,x index=0,y index=1,
  ]
{matlab/two_classes3ext1.csv}
node [pos=0.15,rotate=90,anchor=north] {$\mu=3/8$};
\addplot[
  line width=.3mm,color=green!50!black, dotted,
%  mark=square*,mark repeat=50,mark phase=0,mark size=.5mm,mark options={solid}
  ]
table[
  header=false,x index=0,y index=1,
  ]
{matlab/two_classes3ext2.csv};
\node[point,green!50!black] (pc) at (axis cs: 4,1.25) {};
\node[green!50!black,align=center] at (axis cs: 3.7,1.3) {$\left(\frac{3}{4}, 0\right)$}; %{$\mu_{(1)}=3/4,$\\ $\mu_{(2)}=0$};
\node[point,green!50!black] (pd) at (axis cs: 2.75,1.5) {};
\node[green!50!black,align=center] at (axis cs: 3.05,1.7) {$\left(\frac{1}{2}, \frac{1}{4}\right)$}; %{$\mu_{(1)}=3/8,$\\ $\mu_{(2)}=3/8$};
\node[point,green!50!black] (pe) at (axis cs: 2.625,2.625) {};
\node[green!50!black,align=center] at (axis cs: 3, 2.625) {$\left(\frac{3}{8}, \frac{3}{8}\right)$}; %{$\mu_{(1)}=0,$\\ $\mu_{(2)}=3/4$};
%\addlegendentry{$\mu=3/8$};
%
%
\addplot[
  line width=.3mm,color=black, dotted,
%  mark=square*,mark repeat=50,mark phase=0,mark size=.5mm,mark options={solid}
  ]
table[
  header=false,x index=0,y index=1,
  ]
{matlab/two_classes4ext.csv}
node [pos=0.075,rotate=90,anchor=north] {$\mu=1/2$};
\node[point,black] (pf) at (axis cs: 4,1) {};
\node[black,align=center] at (axis cs: 4,0.8) {$\left(1, 0\right)$}; %{$\mu_{(1)}=1,$\\ $\mu_{(2)}=0$};
\node[point,black] (pg) at (axis cs: 1.5,1.5) {};
\node[black,align=center] at (axis cs: 1.85,1.7) {$\left(\frac{1}{2}, \frac{1}{2}\right)$}; %{$\mu_{(1)}=1/2,$\\ $\mu_{(2)}=1/2$};
%\addlegendentry{$\mu=1/2$};
%
%
\addplot[
  line width=.3mm,color=blue, solid,
%  mark=square*,mark repeat=50,mark phase=0,mark size=.5mm,mark options={solid}
  ]
table[
  header=false,x index=0,y index=1,
  ]
{matlab/two_classes6ext.csv}
node [pos=0.07,rotate=90,anchor=south] {$\mu=3/4$};
\node[point,blue] (ph) at (axis cs: 1.5,1) {};
\node[blue,align=center] at (axis cs: 1.5,0.8) {$\left(1,\frac{1}{2}\right)$}; %{$\mu_{(1)}=1,$\\ $\mu_{(2)}=1/2$};
\node[point,blue] (pi) at (axis cs: 1.25,1.25) {};
\node[blue,align=center] at (axis cs: 0.9,1.25) {$\left(\frac{3}{4}, \frac{3}{4}\right)$}; %{$\mu_{(1)}=3/4,$\\ $\mu_{(2)}=3/4$};
%\addlegendentry{$\mu=3/4$};
%
%
\end{axis}
\end{tikzpicture}
\vspace{-4mm}
\caption{Slice of the NDT region as a function of the fractional cache capacity $\mu$ ($r=1/5$, $J_1=J_2$). The labels indicate the cache allocations $(\mu_{(1)},\mu_{(2)})$ that are required at specific points. The line styles indicate the strategy to be used, with the dashed line corresponding to $\dia$, the dotted lines to $\dib$, and the solid line to $\dic$.} \label{fig:ndtregion}
\end{figure}

The slice of the NDT region at hand is illustrated in Figure~\ref{fig:ndtregion} for $r=1/5$, $J_1=J_2$ and various values of $\mu$. The figure also indicates the values of the cache allocations $(\mu_{(1)},\mu_{(2)})$ that are required to obtain various points on the boundary of the region as well as the delivery strategy that should be used at various segments of the boundary. As it can be seen, the slice of the NDT region is a polyhedron and each linear portion of the boundary corresponds to a different delivery strategy as indicated in the figure (see Sec. III for a correspondence between strategies and NDT tuples $\delta^{(\mathrm{in},\ell)}$).  For instance, it is seen that, for $\mu=3/8$, one has to use a different strategy depending on the operating point: as $\delta_{(1),(2)}$ increases, one needs to switch between the strategies that achieve the NDTs $\dia$ and $\dib$.

Finally, we consider individual latency constraints for different files under a given popularity profile. To this end, we let $a$ and $1-a$ denote the probabilities that a file is requested from class $\mathcal{F}_{1}$ and from class $\mathcal{F}_{2}$, respectively. We then have that the probability $p_{11}$ that two files from class $\mathcal{F}_{1}$ are selected is $p_{11}=a^2$; the probability $p_{12}$ that one file from each class is requested is $p_{12}=2a(1-a)$; and the probability $p_{22}$ that two files from class $\mathcal{F}_{2}$ are requested is $p_{22}=(1-a)^2$. The average latency for a file from a given class is:
\begin{align}
\bar\delta_{(1)} = \mathbb{E}[\delta_{(1)}] &= \frac{p_{11}}{p_{11}+p_{12}}\delta_{(1),(1)} +  \frac{p_{12}}{p_{11}+p_{12}}\delta_{(1),(2)}, \\
\bar\delta_{(2)} = \mathbb{E}[\delta_{(2)}] &= \frac{p_{22}}{p_{22}+p_{12}}\delta_{(1),(1)} +  \frac{p_{12}}{p_{22}+p_{12}}\delta_{(1),(2)}.
\end{align}
Note that $\bar\delta_{(i)}$ is the average latency for files of class $\mathcal{F}_{(i)}$ when averaged over the second requested file. 

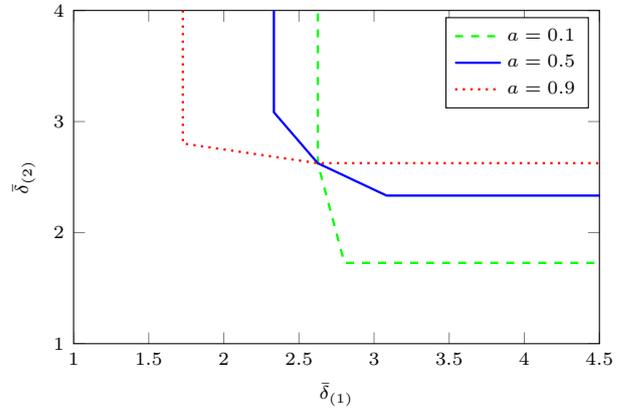
\begin{figure}
\centering
\begin{tikzpicture}
\tikzstyle{point}=[draw,fill,circle,minimum size=3pt,inner sep=0pt]
%\pgfplotsset{every axis legend/.append style={
%        at={(0.5,1.03)},
%        anchor=south}}
\begin{axis}[
	height=60mm,
	width=85mm,
	xlabel=$\bar\delta_{(1)}$, %$\mathbb{E}[\delta_{(1)}]$,
	ylabel=$\bar\delta_{(2)}$,
	xmin=1,
	xmax=4.5,
	ymin=1,
	ymax=4,
	font=\scriptsize,
]

\addplot[
  line width=.3mm,color=green, dashed,
  ]
table[
  header=false,x index=0,y index=1,
  ]
{matlab/two_classes_average_1.csv};
\addlegendentry{$a=0.1$};

\addplot[
  line width=.3mm,color=blue, solid,
  ]
table[
  header=false,x index=0,y index=1,
  ]
{matlab/two_classes_average_5.csv};
\addlegendentry{$a=0.5$};

\addplot[
  line width=.3mm,color=red, dotted,
  ]
table[
  header=false,x index=0,y index=1,
  ]
{matlab/two_classes_average_9.csv};
\addlegendentry{$a=0.9$};

\end{axis}
\end{tikzpicture}
\vspace{-4mm}
\caption{Optimal trade-off between the average delivery latencies for the files of two classes for different popularity profiles defined by $a$ ($\mu=3/8$,$r=1/5$, $J_1=J_2$).}
\label{fig:average}
\end{figure}

In Figure~\ref{fig:average}, we illustrate the optimal trade-off between the average NDTs $\bar\delta_{(1)}$ and $\bar\delta_{(2)}$ that arises from adjusting the cache allocations among the two classes. In the figure we have set $\mu=3/8$, $r=1/5$, $J_1=J_2$ and considered various values of $a$. The figure confirms that obtaining lower average delivery latencies for some files entails a larger average delivery latencies for other files due to the limited cache capacities.

%\tikzexternaldisable
%%%%%%%%%%%%%%%%%%%%%%%%%%%%%%%%%%%%%%%%%%%%%%%%%%%%%%
%
%%%%%%%%%%%%%%%%%%%%%%%%%%%%%%%%%%%%%%%%%%%%%%%%%%%%%%
%%%%%%%%%%%%%%%%%%%%%%%%%%%%%%%%%%%%%%%%%%%%%%%%%%%%%%
%
%%%%%%%%%%%%%%%%%%%%%%%%%%%%%%%%%%%%%%%%%%%%%%%%%%%%%%
%\tikzexternalenable

%\tikzexternaldisable
%%%%%%%%%%%%%%%%%%%%%%%%%%%%%%%%%%%%%%%%%%%%%%%%%%%%%%
%
%%%%%%%%%%%%%%%%%%%%%%%%%%%%%%%%%%%%%%%%%%%%%%%%%%%%%%
%%%%%%%%%%%%%%%%%%%%%%%%%%%%%%%%%%%%%%%%%%%%%%%%%%%%%%
%
%%%%%%%%%%%%%%%%%%%%%%%%%%%%%%%%%%%%%%%%%%%%%%%%%%%%%%
%\tikzexternalenable

%%%%%%%%%%%%%%%%%%%%%%%%%%%%%%%%%%%%%%%%%%%%%%%%%%%%%%
%
%
%
%%%%%%%%%%%%%%%%%%%%%%%%%%%%%%%%%%%%%%%%%%%%%%%%%%%%%%
\section{Conclusions}
This work characterized the set of delivery latencies supported by an F-RAN with two ENs and two users in the high SNR regime, when allowing for any cache partition across the files in a set of popular contents. Various aspects call for further investigation, including the explicit minimization of the average delivery latency as a function of the content popularity profile, the extension of the main results to any number of ENs and users (see~\cite{sengupta2016cloud} for the case of uniform file popularity) and the derivation of an extended NDT region in which the same contents may be requested by multiple users.

%%%%%%%%%%%%%%%%%%%%%%%%%%%%%%%%%%%%%%%%%%%%%%%%%%%%%%
%
%
%
%%%%%%%%%%%%%%%%%%%%%%%%%%%%%%%%%%%%%%%%%%%%%%%%%%%%%%

%%%%%%%%%%%%%%%%%%%%%%%%%%%%%%%%%%%%%%%%%%%%%%%%%%%%%%
%
%
%
%%%%%%%%%%%%%%%%%%%%%%%%%%%%%%%%%%%%%%%%%%%%%%%%%%%%%%
\bibliographystyle{IEEEtran}
\bibliography{IEEEabrv,fogran_nonuniform}

%%%%%%%%%%%%%%%%%%%%%%%%%%%%%%%%%%%%%%%%%%%%%%%%%%%%%%
%
%
%
%%%%%%%%%%%%%%%%%%%%%%%%%%%%%%%%%%%%%%%%%%%%%%%%%%%%%%
\appendices

%%%%%%%%%%%%%%%%%%%%%%%%%%%%%%%%%%%%%%%%%%%%%%%%%%%%%%
%
%
%
%%%%%%%%%%%%%%%%%%%%%%%%%%%%%%%%%%%%%%%%%%%%%%%%%%%%%%
\section{Proof of Lemma~\ref{lem:convex}} \label{app:convex}
The proof is based on cache sharing and file splitting, similar to~\cite{sengupta2016cloud}.

Let $\delta$ and $\delta'$ be NDT tuples that are achievable with policies $\pi$ and $\pi'$ and cache partitions $\bm{\mu}$ and $\bm{\mu'}$, respectively, that satisfy~\eqref{eq:capacityconstraint}. Moreover, let $0<\alpha<1$ be arbitrary. We will prove that the region $\Delta^*(\mu, r)$ is convex by demonstrating that the NDT tuple $\alpha \delta+(1-\alpha)\delta'$ (with entry-wise sum) is achievable for a cache partition $\alpha\bm{\mu}+(1-\alpha)\bm{\mu'}$, which also satisfies~\eqref{eq:capacityconstraint}.

We start by splitting each of the files in $\mathcal{F}$ in two parts of sizes $\alpha L$ and $(1-\alpha)L$ bits. Moreover, we split the cache at each EN in two parts of sizes $\alpha\mu JL$ bits and $(1-\alpha)\mu JL$ bits. Now, to achieve the NDT tuple $\alpha \delta+(1-\alpha)\delta'$ we transmit the first fraction of the pair of requested files using policy $\pi$ and the first part of the caches and then transmit the second part of the files using $\pi'$ and the second part of the caches. This policy uses cache partition $\alpha\bm{\mu}+(1-\alpha)\bm{\mu'}$ and hence satisfies the cache capacity constraints~\eqref{eq:capacityconstraint}. Furthermore, it achieves the desired NDT tuple $\alpha \delta+(1-\alpha)\delta'$, since the NDT is proportional to the file size.

It remains to be shown that for any $\bm{\mu}$, $\Delta^*(\bm{\mu},r)$ is convex. This follows directly by taking in the above argument $\bm{\mu'}=\bm{\mu}$ and observing that the overall cache allocation in the two phase policy achieving $\alpha \delta+(1-\alpha)\delta'$ is again equal to $\bm{\mu}$.

%%%%%%%%%%%%%%%%%%%%%%%%%%%%%%%%%%%%%%%%%%%%%%%%%%%%%%
%
%
%
%%%%%%%%%%%%%%%%%%%%%%%%%%%%%%%%%%%%%%%%%%%%%%%%%%%%%%
\section{Proof of Theorem~\ref{th:inner}} \label{app:inner}
We analyze the fronthaul and edge NDTs using Lemma~\ref{lem:existing}. W.l.o.g.\ we consider $\mu_i\leq \mu_j.$ We focus on the case $r\leq 1$ since the result for the case $r>1$ is an immediate consequence of Lemma~\ref{lem:existing}.

First, for $\mu_i<\frac{1}{2}$ and $\mu_j<\frac{1}{2}$, the HT delivery of bits $\mu_iL,\dots\mu_jL$ and $(1-\mu_j)L,\dots,(1-\mu_i)L$ of file $F_i$ to EN$_1$ and EN$_2$, respectively, is done over parallel channels to EN$_1$ and EN$_2$ and takes a fronthaul NDT $\delta^f = \frac{1}{r}\left(\mu_j-\mu_i\right)$ by Lemma \ref{lem:existing}. In the X-IA phase we deliver four parts, two of each file, of $\mu_jL$ bits, and hence the edge NDT is $\delta^e=3 \mu_j.$ The fronthaul and edge NDTs to deliver the remaining $(1-2\mu_j)L$ bits of both files through ST+ZF are $\delta^f=\frac{1}{r}\left(1-2\mu_j\right)$ and $\delta^e=1-2\mu_j$, respectively. Summing all the NDT terms gives the result.

Second, for $\mu_i\leq\frac{1}{2}$ and  $\mu_j\geq\frac{1}{2}$, the HT transmission of $(1/2-\mu_i)L$ bits to each EN requires a fronthaul NDT $\delta^f=\frac{1}{r}(1/2-\mu_i)$. The X-IA phase, in which messages of $L/2$ bits are transmitted, instead takes an edge NDT of $\delta^e=3/2$.

Third, for $\mu_i>\frac{1}{2}$ and $\mu_j>\frac{1}{2}$, the X-IA phase delivers four messages of $(1-\mu_i)L$ bits using an edge NDT $\delta^e=3(1-\mu_i)$. The ZF phase instead delivers $(2\mu_i-1)L$ bits of each file and requires an edge NDT $\delta^e=2\mu_i-1.$

%%%%%%%%%%%%%%%%%%%%%%%%%%%%%%%%%%%%%%%%%%%%%%%%%%%%%%
%
%
%
%%%%%%%%%%%%%%%%%%%%%%%%%%%%%%%%%%%%%%%%%%%%%%%%%%%%%%
\section{Proof of Theorem~\ref{th:outer}} \label{app:outer}
In this appendix, we let $\epsilon_P$ denote any function such that $\epsilon_P/\log(P)\to 0$ as $P\to\infty$ and let $\epsilon_L$ denote any function such that $\epsilon_L\to 0$ as $L\to\infty$. Furthermore, we drop the dependence of $n^e_D$ and $n^f_D$ on $D$ in order to streamline the notation.

We start with two technical results that appear in~\cite{sengupta2016cloud}.
\begin{lemma}[\!\!\cite{sengupta2016cloud}, Lemma~6] \label{lem:senguptaA}
For $k=1,2$ we have
\begin{equation}
H(F_i, F_j \given Y_1, S_k, F_{[1:J]\setminus\{i, j\}}) \leq L\epsilon_L + n^e\epsilon_P.
\end{equation}
\end{lemma}

\begin{lemma}[\!\!\cite{sengupta2016cloud}, Lemma 5]  \label{lem:senguptaB}
For $k=1,2$ we have
\begin{equation}
I(F_i, F_j\with Y_k\given F_{[1:J]\setminus\{i,j\}}) \leq n^e\log P + n^e\epsilon_P.
\end{equation}
\end{lemma}
Note that Lemma~\ref{lem:senguptaB} does not appear in this form in~\cite{sengupta2016cloud}, but it follows directly from Lemma 5 in~\cite{sengupta2016cloud}.

We will now develop several bounds on linear combinations of the edge NDT $\delta^e_{i,j}$ and fronthaul NDT $\delta^f_{i,j}$ for any sequence of feasible schemes. These bounds will be used to construct a bound on the NDT $\delta_{i,j}=\delta^e_{i,j}+\delta^f_{i,j}$. The following three lemmas provide such bounds.

\begin{lemma} \label{lem:converseA}
Let $i,j\in[1:J]$. Then, any sequence of achievable strategies satisfies the inequality
\begin{equation}
\delta_{i,j}^e + r\delta_{i,j}^f  \geq 2 - \min\{\mu_{1,i}, \mu_{2,i}, \mu_{1,j}, \mu_{2,j}\}.
\end{equation}
%for all $m=1,2$ and all $k=i,j$.
\end{lemma}
\begin{IEEEproof}
W.l.o.g.\ we prove the inequality $\delta_{i,j}^e + r\delta_{i,j}^f \geq 2 - \mu_{m,j}$, $m=1,2.$ First, we can write
 \begin{align}
 2L =&\ H(F_i, F_j\given F_{[1:J]\setminus\{i, j\}}) \\
 =&\ I( F_{i}, F_{j}\with Y_1, S_m, U_m\given F_{[1:J]\setminus\{i, j\}}) \notag \\
 &+ H(F_i, F_j\given Y_1, S_m, U_m, F_{[1:J]\setminus\{i, j\}}) \\
% \leq&\ I( F_i, F_j\with Y_1, S_m, U_m\given F_{[1:J]\setminus\{i,j\}})
% 	+ L\epsilon_L \notag \\
%	&+ H(F_j \given Y_1, S_m, U_m, F_{[1:J]\setminus\{j\}}) \\
\leq&\ I( F_i, F_j\with Y_1, S_m, U_m\given F_{[1:J]\setminus\{i, j\}})
 	+ L\epsilon_L \notag \\
	&+ n^e\epsilon_P \label{eq:lemConverseA5},
 \end{align}
where the equality follows from the independence of files
 and the
inequality follows from Lemma~\ref{lem:senguptaA} and the fact that conditioning on $U_m$ reduces entropy. Next, we write
\begin{align}
I( F_i, & F_j\with Y_1, S_m, U_m\given F_{[1:J]\setminus\{i,j\}}) \notag \\
\leq&\ I( F_i, F_j\with Y_1, F_i, S_m, U_m\given F_{[1:J]\setminus\{i,j\}}) \\
=&\ I(F_i, F_j\with Y_1\given F_{[1:J]\setminus\{i,j\}}) \notag \\ %\\
	&+ I( F_i, F_j\with F_i\given F_{[1:J]\setminus\{i,j\}}, Y_1) \notag \\
	&+ I( F_i, F_j\with S_m, U_m\given F_{[1:J]\setminus\{j\}}, Y_1)  \label{eq:lemConverseA1} \\
\leq&\ n^e\log P + n^e\epsilon_P + L\epsilon_L \notag \\
&+ H(S_{m}, U_m\given F_{[1:J]\setminus\{j\}}) \label{eq:lemConverseA2} \\
\leq&\ n^e\log P + n^e\epsilon_P + L\epsilon_L + n^f r \log P \notag \\
&+ H(S_{m}\given F_{[1:J]\setminus\{j\}}) \label{eq:lemConverseA3} \\
\leq&\ n^e\log P + n^e\epsilon_P + L\epsilon_L + n^f r \log P + \mu_{m,j} L, \label{eq:lemConverseA4}
\end{align}
where: \eqref{eq:lemConverseA2} follows by bounding the first term in~\eqref{eq:lemConverseA1} using Lemma~\ref{lem:senguptaB} and the second term in~\eqref{eq:lemConverseA1} using Fano's inequality; \eqref{eq:lemConverseA2} follows from $H(U_m)\leq n^f r \log P$ by the fronthaul capacity constraints; and \eqref{eq:lemConverseA4} follows from the definition of $\mu_{m,j}$.

Combining~\eqref{eq:lemConverseA5} and~\eqref{eq:lemConverseA4} and rewriting in terms of $\delta^e$ and $\delta^f$ gives
\begin{equation}
\delta^e\left(1 + \frac{\epsilon_P}{\log P}\right) + \delta^f r \geq 2  - \mu_{m,j} - \epsilon_L,
\end{equation}
and the result follows by taking $L\to\infty$ and $P\to\infty$.
\end{IEEEproof}

\begin{lemma} \label{lem:converseB}
Let $i,j\in[1:J]$. Then, any sequence of achievable strategy satisfies the inequality
\begin{equation}
2r\delta_{i,j}^f \geq 1 - \min\{\mu_{1,i}+\mu_{2,i}, \mu_{1,j}+\mu_{2,j}\}.
\end{equation}
%for all $k=i,j$.
\end{lemma}
\begin{IEEEproof}
W.l.o.g.\ we prove the inequality $2r\delta_{i,j}^f \geq 1 - \mu_{1,i} - \mu_{2,i}$. We have
\begin{align}
 L =&\
% H(F_i) \\
% &= H(F_i \given F_{[1:J]\setminus\{i\}}) \\
% &=
 I( F_i \with S_1, U_1, S_2, U_2\given F_{[1:J]\setminus\{i\}}) \notag \\
 &+ H(F_i \given S_1, U_1, S_2, U_2, F_{[1:J]\setminus\{i\}}) \\
 \leq&\ H(S_1, S_2 \given F_{[1:J]\setminus\{i\}}) \notag \\
 &+ H(U_1\given F_{[1:J]\setminus\{i\}}) + H(U_2\given F_{[1:J]\setminus\{i\}})
 	+ L\epsilon_L \\
\leq&\ \mu_{1,i} + \mu_{2,i} + 2rn^f\log P + L\epsilon_L,
 \end{align}
where the first inequality follows from the equality $H(S_1, U_1^{T_F}, S_2, U_2^{T_F} \given F_{[1:J]})=0$ and from Fano's inequality, since file $F_i$ can be recovered from $S_1, U_1^{T_F}, S_2, U_2^{T_F}$ given that the input signals are functions of these variables. The second inequality follows from the fronthaul rate constraint. The result follows by taking the limits $L\to\infty$ and $P\to\infty$.
\end{IEEEproof}

\begin{lemma} \label{lem:converseC}
Let $i,j\in[1:J]$. Then, any sequence of achievable strategy satisfies the inequality
%For all $i,j=1, \dots, N$ we have $
\begin{equation}
2r\delta^f \geq 2 - \mu_{1,i} - \mu_{2,i} - \mu_{1,j} - \mu_{2,j}.
\end{equation}
\end{lemma}
\begin{IEEEproof}
We have
 \begin{align}
 2L =&\
% H(F_i) \\
% &= H(F_i \given F_{[1:J]\setminus\{i\}}) \\
% &=
 I( F_i, F_j \with S_1, U_1, S_2, U_2\given F_{[1:J]\setminus\{i,j\}}) \notag \\
 &+ H(F_i, F_j \given S_1, U_1, S_2, U_2, F_{[1:J]\setminus\{i,j\}}) \\
 \leq&\ H(S_1, S_2 \given F_{[1:J]\setminus\{i,j\}}) + H(U_1\given F_{[1:J]\setminus\{i,j\}}) \notag \\
  &+ H(U_2\given F_{[1:J]\setminus\{i,j\}})
 	+ L\epsilon_L \\
\leq&\ \mu_{1,i} + \mu_{2,i} + \mu_{1,j} + \mu_{2,j} + 2rn^f\log P + L\epsilon_L,
 \end{align}
where the first inequality follows from the equality $H(S_1, U_1, S_2, U_2 \given F_{[1:J]})=0$ and from Fano's inequality, since the files $F_i$ and $F_j$ can be recovered from the variables $S_1, U_1, S_2, U_2$ as discussed above.
\end{IEEEproof}

We now summarize the constraints of Lemmas~\ref{lem:converseA}--\ref{lem:converseC} as
\begin{align}
\delta_{i,j}^e + r\delta_{i,j}^f \geq&\ 2 - \min\{\mu_{1,i}, \mu_{2,i}, \mu_{1,j}, \mu_{2,j}\}, \label{eq:linearcomb1} \\
r\delta_{i,j}^f \geq&\ \frac{1}{2} - \frac{1}{2}\min\{\mu_{1,i} + \mu_{2,i}, \mu_{1,j} + \mu_{2,j}\}, \label{eq:linearcomb2}\\
r\delta_{i,j}^f \geq&\ 1 - \frac{\mu_{1,i} + \mu_{2,i}}{2} - \frac{\mu_{1,j}+ \mu_{2,j}}{2}, \label{eq:linearcomb3} \\
\intertext{and we add}
\delta_{i,j}^e \geq&\ 1, \label{eq:linearcomb4}\\
\delta_{i,j}^f \geq&\ 0, \label{eq:linearcomb5}
\end{align}
where~\eqref{eq:linearcomb4} holds since the edge NDT in an interference-free wireless channel is $1$. The required bounds for Theorem~\ref{th:outer} are finally obtained by taking various linear combinations of~\eqref{eq:linearcomb1}--\eqref{eq:linearcomb5}. In particular, for the case $r\leq 1$, by taking~\eqref{eq:linearcomb1} and~\eqref{eq:linearcomb3} with weights~$1$ and $(1/r-1)$, respectively, we obtain
\begin{multline}
\left[  \delta_{i,j}^e + r\delta_{i,j}^f \right] + \left(\frac{1}{r}-1\right)\left[ r\delta_{i,j}^f \right] \geq \\
\left[  2 - \min\{\mu_{1,i}, \mu_{2,i}, \mu_{1,j}, \mu_{2,j}\}\right] \\
+ \left(\frac{1}{r}-1\right)\left[ 1 - \frac{\mu_{1,i} + \mu_{2,i}}{2} - \frac{\mu_{1,j} + \mu_{2,j}}{2} \right],
\end{multline}
from which it follows that $\delta_{i,j}\geq\doa$.

Next, for $r\leq 1$, taking~\eqref{eq:linearcomb1} and~\eqref{eq:linearcomb2} with weights~$1$ and $(1/r-1)$, respectively, gives

\begin{multline}
\left[  \delta_{i,j}^e + r\delta_{i,j}^f \right] + \left(\frac{1}{r}-1\right)\left[ r\delta_{i,j}^f \right] \geq \\
 \left[  2 - \min\{\mu_{1,i}, \mu_{2,i}, \mu_{1,j}, \mu_{2,j}\}\right] \\
+ \left(\frac{1}{r}-1\right)\left[ \frac{1}{2} - \frac{1}{2}\min\{\mu_{1,i} + \mu_{2,i}, \mu_{1,j}+ \mu_{2,j}\} \right],
\end{multline}
which is equivalent to $\delta_{i,j}\geq\dob$.

Continuing with $r\leq 1$, the inequality $\delta_{i,j}\geq\doc$ follows from
\begin{multline}
\left[  \delta_{i,j}^e + r\delta_{i,j}^f \right] + \left(\frac{1}{r}-1\right)\left[ r\delta_{i,j}^f \right] \geq \\
\left[  2 - \min\{\mu_{1,i}, \mu_{2,i}, \mu_{1,j}, \mu_{2,j}\}\right] + \left(\frac{1}{r}-1\right)\cdot 0,
\end{multline}
which is obtained by taking~\eqref{eq:linearcomb1} and~\eqref{eq:linearcomb5} with weights~$1$ and $(1/r-1)$, respectively.

Finally, for $r>1$, by taking~\eqref{eq:linearcomb1} and~\eqref{eq:linearcomb4} with weights~$1$ and $(1-1/r)$, respectively, we obtain
\begin{multline}
\frac{1}{r}\left[  \delta_{i,j}^e + r\delta_{i,j}^f \right] + \left(1 - \frac{1}{r}\right)\left[ \delta_{i,j}^e \right] \geq \\
 \frac{1}{r}\left[  2 - \min\{\mu_{1,i}, \mu_{2,i}, \mu_{1,j}, \mu_{2,j}\}\right] + \left(1-\frac{1}{r}\right)\cdot 1,
\end{multline}
from which it follows that $\delta_{i,j}\geq\dod$.
%

%%%%%%%%%%%%%%%%%%%%%%%%%%%%%%%%%%%%%%%%%%%%%%%%%%%%%%
%
%
%
%%%%%%%%%%%%%%%%%%%%%%%%%%%%%%%%%%%%%%%%%%%%%%%%%%%%%%
\section{Proof of Theorem~\ref{th:region}} \label{app:region}
To prove Theorem 3, we need to show that the NDT region \eqref{eq:ndtasunion} is given as 
\begin{equation} \label{eq:appregiontoshow}
\Delta^*(\mu,r) = \bigcup_{\bm{\mu}\in\mathcal{U}} \Dinner(\bm{\mu}, r),
\end{equation}
where we have introduced the set
\begin{equation}
\mathcal{U} = \left\{ \bm{\mu}\ \middle|\ \forall i\in[1:J]: \mu_{1,i}=\mu_{2,i}=\mu_i, \text{ and } \sum_{j=1}^J\mu_{j}\leq\mu  \right\}
\end{equation} for convenience of notation. The proof consists of two parts. 1) We demonstrate that for any $\bm{\tilde\mu}\not\in\mathcal{U}$ such that \eqref{eq:capacityconstraint} holds, there exists a $\bm{\hat\mu}\in\mathcal{U}$ for which
\begin{equation} \label{eq:appregion1}
\Delta^*(\bm{\tilde\mu}, r) \subset \Delta^*(\bm{\hat\mu}, r)
\end{equation}
and \eqref{eq:capacityconstraint} hold. This allows us to restrict the union in ~\eqref{eq:ndtasunion} with no loss of optimality to set $\mathcal{U}$ as
\begin{equation} \label{eq:appregion2}
 \Delta^*(\mu,r) = \bigcup_{\bm{\mu}\in\mathcal{U}} \Delta^*(\bm{\mu}, r).
\end{equation}
2) We show that, for any $\bm{\mu}\in\mathcal{U}$, we have
\begin{equation} \label{eq:appregion3}
\Delta^*(\bm{\mu}, r) = \Dinner(\bm{\mu}, r) = \Douter(\bm{\mu}, r),
\end{equation}
which reduces~\eqref{eq:appregion2} to~\eqref{eq:appregiontoshow}, hence concluding the proof. Details are provided next.

\enlargethispage{-60mm}

1) To prove~\eqref{eq:appregion1}, we start by constructing the mentioned cache partition $\bm{\hat\mu}$ from $\bm{\tilde\mu}\not\in\mathcal{U}$ as
\begin{equation}
\hat\mu_{i} = \hat\mu_{1,i} = \hat\mu_{2,i} = \frac{\tilde\mu_{1,i}+\tilde\mu_{2,i}}{2}.
\end{equation}
The choice $\bm{\hat\mu}$ satisfies the capacity constraint~\eqref{eq:capacityconstraint}, since we have
\begin{equation}
\sum_{i=1}^J \hat\mu_{m,i} = \frac{1}{2}\left(\sum_{i=1}^J\tilde\mu_{1,i} + \sum_{i=1}^J\tilde\mu_{2,i}\right) \leq J\mu,
\end{equation}
where the inequality holds because of the constraints $\sum_{i=1}^J\tilde\mu_{1,i}\leq J\mu$ and $\sum_{i=1}^J\tilde\mu_{2,i}\leq J\mu$. We now argue that, for an arbitrary $(i,j)\in\mathcal{D}$, the achievable NDT under $\bm{\hat\mu}$ by Theorem 1 is strictly smaller than the lower bound on the NDT under $\bm{\tilde\mu}$ obtained in Theorem 2, \ie
\begin{equation}
\dii(\bm{\hat\mu}, r) < \doo(\bm{\tilde\mu}, r).
\end{equation} This would conclude the proof of part 1). To this end, we leverage the following inequalities: for all $\ell=[1:4]$, we have
\begin{equation} \label{eq:appregionA}
\dol(\bm{\hat\mu}, r) \leq \dol(\bm{\tilde\mu}, r),
\end{equation}
with equality if and only if $\tilde\mu_{1,i}=\tilde\mu_{2,i}=\hat\mu_i$ and $\tilde\mu_{1,j}=\tilde\mu_{2,j}=\hat\mu_j$; and
\begin{equation} \label{eq:appregionB}
\dil(\bm{\mu}, r) = \dol(\bm{\mu}, r),
\end{equation}
for all $\bm{\mu}\in\mathcal{U}.$ The proofs of~\eqref{eq:appregionA} and~\eqref{eq:appregionB} are deferred to the end of this appendix. Now, w.l.o.g.\ assume that $\dii(\bm{\hat\mu}, r)=\dil(\bm{\hat\mu}, r)$. Then, we can write
\begin{multline}
\dii(\bm{\hat\mu}, r)
= \dil(\bm{\hat\mu}, r)
= \dol(\bm{\hat\mu}, r) \\
< \dol(\bm{\tilde\mu}, r)
\leq \doo(\bm{\tilde\mu}, r),
\end{multline}
where the second equality follows from~\eqref{eq:appregionB}, the first inequality from~\eqref{eq:appregionB} and the second inequality from the definition of $\doo(\bm{\tilde\mu}, r)$. This proves~\eqref{eq:appregion1}.

2) Equation~\eqref{eq:appregion3} follows directly from~\eqref{eq:appregionB}, since the latter equality shows that any point on the boundary of the region $\Douter(\bm{\tilde\mu}, r)$ can be achieved with the scheme proposed in Theorem 1.

\emph{Proof of~\eqref{eq:appregionA} and~\eqref{eq:appregionB}}: Observe that, for all $\ell\in[1:4]$, we have
\begin{multline}
\dol(\bm{\tilde\mu}, r) - \dol(\bm{\hat\mu}, r) \\
= \min\left\{\hat\mu_{1,i},\hat\mu_{2,i}, \hat\mu_{1,j}, \hat\mu_{2,j}\right\} - \min\left\{\tilde\mu_{1,i},\tilde\mu_{2,i}, \tilde\mu_{1,j}, \tilde\mu_{2,j}\right\} \\
= \min\left\{\frac{\tilde\mu_{1,i}+\tilde\mu_{2,i}}{2}, \frac{\tilde\mu_{1,j}+\tilde\mu_{2,j}}{2}\right\} \\
- \min\left\{\min\{\tilde\mu_{1,i},\tilde\mu_{2,i}\}, \min\{\tilde\mu_{1,j}, \tilde\mu_{2,j}\}\right\}.
\end{multline}
Now,~\eqref{eq:appregionA} follows from the inequality
\begin{equation}
\frac{\tilde\mu_{1,k}+\tilde\mu_{2,k}}{2} \geq \min\left\{\tilde\mu_{1,k},\tilde\mu_{2,k}\right\},
\end{equation}
for $k=1,2$, which holds with equality if and only if $\tilde\mu_{1,k}=\tilde\mu_{2,k}$.

Finally, for~\eqref{eq:appregionB}, with $\bm{\mu}\in\mathcal{C}$, we have
\begin{multline}
\doa(\bm{\mu}, r) = 1 + \frac{1}{r} - \min\left\{\mu_{i}, \mu_{j}\right\} \\
-\left(\frac{1}{r}-1\right)\left(  \min\left\{\mu_{i}, \mu_{j}\right\} + \max\left\{\mu_{i}, \mu_{j}\right\}  \right)
= \dia(\bm{\mu}, r).
\end{multline}
In a similar fashion, it follows that $\dil(\bm{\mu}, r)=\dol(\bm{\mu}, r)$ for all $\ell=[1:4]$.

\end{document}